\documentclass[conference,letterpaper]{IEEEtran}
\IEEEoverridecommandlockouts
\usepackage{tikz}
\usepackage[section]{placeins}
\usepackage{bm}
\usepackage{lettrine}
\usepackage{bbm}
\usepackage{subcaption}
\usepackage{float}
\usepackage{graphicx}
\usepackage{epstopdf}
\usepackage{mathtools}
\usepackage{algorithmic}
\usepackage{algorithm}
\usepackage{framed}
\usepackage{dsfont}
\usepackage{overpic}
\usepackage{amsfonts}
\usepackage{amssymb}
\usepackage{amsmath}
\usepackage{latexsym}
\usepackage{times}

\allowdisplaybreaks
\makeatletter
\def\BState{\State\hskip-\ALG@thistlm}
\makeatother

\DeclareMathOperator*{\argmax}{arg\,max}
\DeclareRobustCommand*{\IEEEauthorrefmark}[1]{%
	\raisebox{0pt}[0pt][0pt]{\textsuperscript{\footnotesize\ensuremath{#1}}}}
\newcommand{\inprod}[2]{\left\langle #1,#2 \right\rangle}
\newcommand{\bvec}[1]{\bm{#1}}
\newcommand{\real}{\mathbb{R}}

\newcommand{\opnorm}[2]{\left\|#1\right\|_{#2}}
\newcommand{\abs}[1]{\left|#1\right|}

\newtheorem{theorem}{Theorem}[section]

\newtheorem{proposition}[theorem]{Proposition}

\newcommand{\qed}{\nobreak \ifvmode \relax \else
      \ifdim\lastskip<1.5em \hskip-\lastskip
      \hskip1.5em plus0em minus0.5em \fi \nobreak
      \vrule height0.75em width0.5em depth0.25em\fi}
\newcommand{\prob}[1]{\mathbb{P}\left(#1\right)}
\newcommand{\expect}[1]{\mathbb{E}\left[#1\right]}
\newcommand{\expectsuff}[2]{\mathbb{E}_{#1}\left[#2\right]}

\newcommand{\indicator}[1]{\mathbbm{1}_{\left\{#1\right\}}}

\newtheorem{lem}{Lemma}[section]
\newtheorem{thm}{Theorem}[section]

\begin{document}
\title{Online Caching with Optimal Switching Regret}
\author{%
\IEEEauthorblockN{Samrat Mukhopadhyay\IEEEauthorrefmark{1}, Abhishek Sinha\IEEEauthorrefmark{2}}
  \IEEEauthorblockA{Dept. of Electrical Engineering, Indian Institute of Technology Madras\\
                    Chennai 600036, India\\ 
                    }
  Email:  
   \IEEEauthorrefmark{1} samratphysics@gmail.com, 
  \IEEEauthorrefmark{2} abhishek.sinha@ee.iitm.ac.in\\

}
        

\maketitle
\begin{abstract}
We consider the classical uncoded caching problem from an online learning point-of-view. A cache of limited storage capacity can hold $C$ files at a time from a large catalog. A user requests an arbitrary file from the catalog at each time slot. Before the file request from the user arrives, a caching policy populates the cache with any $C$ files of its choice. In the case of a cache-hit, the policy receives a unit reward and zero rewards otherwise. In addition to that, there is a cost associated with fetching files to the cache, which we refer to as the \emph{switching cost}. The objective is to design a caching policy that incurs minimal regret while considering both the rewards due to cache-hits and the switching cost due to the file fetches.   
The main contribution of this paper is the switching regret analysis of a \textsc{Follow the Perturbed Leader}-based anytime caching policy, which is shown to have an order optimal switching regret. In this pursuit, we improve the best-known switching regret bound for this problem by a factor of $\Theta(\sqrt{C}).$ We conclude the paper by comparing the performance of different popular caching policies using a publicly available trace from a commercial CDN server.
\end{abstract}	

\section{Introduction}
\lettrine{C}{aching} is a fundamental online optimization problem that has been extensively investigated in the literature by different research communities. \emph{Competitive Ratio}, which quantifies the performance of any online policy by computing the ratio of the cost incurred by an online policy to the cost incurred by a clairvoyant offline optimal policy on the same input sequence, has been the classical performance metric for benchmarking different caching policies \cite{dan1990approximate, lee1999existence, pedarsani2016online}. Policies, such as the Least Recently Used (\textsf{LRU}), Least Frequently Used (\textsf{LFU}), and First-in-First-Out (\textsf{FIFO}) are known to achieve the optimal competitive ratio when the cost function is taken to be the number of cache-misses \cite{albers}. However, since the competitive ratio metric is multiplicative in nature, there could be a large gap in the absolute number of cache-misses between an online policy (even with an excellent competitive ratio) and the optimal offline policy. Consequently, in recent years, a considerable amount of research effort has been dedicated to designing caching policies with more robust (\emph{e.g.,} additive) performance guarantees. In particular, with the advances in the theory of online learning \cite{cesa2006prediction}, several recent papers have investigated the caching problem from the regret-minimization standpoint. Using the framework of Online Convex Optimization (OCO), the papers \cite{bhattacharjee2020fundamental}, \cite{paschos2019learning} take the first step in designing a regret-optimal caching policy without and with coding, respectively. The paper \cite{paria2020caching} considers the regret-optimal uncoded caching problem on a Bipartite network. It should be noted that, in a caching network, repeatedly fetching files to the local caches from remote servers costs considerable latency and bandwidth penalty. 
However, most of the previous works on caching ignore the switching cost in their cost function and focus only on the hit rates. On a parallel line of research, several papers from the online learning community have studied the problem of switching regret minimization in different settings. Examples include the Follow the Lazy Leader (FLL) algorithm proposed by Kalai and Vempala \cite{kalai2005efficient}, the ``Shrinking Dartboard" algorithm by Geulen et al. \cite{geulen2010regret}, and the Prediction by random-walk perturbations algorithm by Devroye et al. \cite{devroye2015random}. However, all of the above algorithms are known to suffer from sub-optimal regret $\tilde{O}(C^{3/2}\sqrt{T})$ in the caching problem (see Table I of \cite{devroye2015random}), where $C$ is the size of the cache, $T$ is the horizon-length, and the fetching cost per file is assumed to be unity.
\paragraph{Contributions} The Multiplicative Weight policy (also known as \emph{Hedge}, or Exponential Weight) is a well-known prediction strategy in the classical experts setting \cite{freund1997decision, littlestone1994weighted}. In a recent paper, Daniely et al. proposed a version of the  multiplicative-weight policy (\textsf{MW}) for the caching problem. They showed that the regret of the \textsf{MW} policy, including the switching cost, is bounded by $\tilde{O}(C\sqrt{T})$ (Theorem $31$ of \cite{daniely2019competitive}). The authors also proposed an efficient implementation of the \textsf{MW} policy in this context, which, otherwise, suffers from an exponential complexity in a na\"ive implementation. However, the resulting caching policy is still computationally intensive as it involves repeated sampling from a recursively defined distribution. Consequently, a natural question in this context is whether there exists a simpler caching policy with a better performance guarantee. In this paper, we affirmatively resolve the above question. Using the \textsf{Follow the Perturbed Leader (FTPL)}-based caching policy proposed in \cite{bhattacharjee2020fundamental}, we show that, in addition to being trivial to implement, the \textsf{FTPL}-based caching policy enjoys a switching regret bound of $\tilde{O}(\sqrt{CT})$. Hence, it improves the regret guarantee of the \textsf{MW} policy by a factor of $\Theta(\sqrt{C}).$ The lower bound established in \cite{bhattacharjee2020fundamental} implies that our regret bound is optimal and can not be improved further. 

Note that we do not assume any stochastic model for the file request sequence, which could be generated by an \emph{oblivious} adversary. There is extensive literature on caching, which assumes some stationary stochastic model for the file request sequence, such as the Independent Reference Model (IRM) \cite{shanmugam2013femtocaching, traverso2013temporal}. However, these models may be a poor choice for non-stationary requests, which is typical for internet traffic with transient content popularity.
\section{System Model}
\label{sec:sys-model}

\begin{figure}[t!]
    \centering
    \includegraphics[height=1.25in, width = 3in]{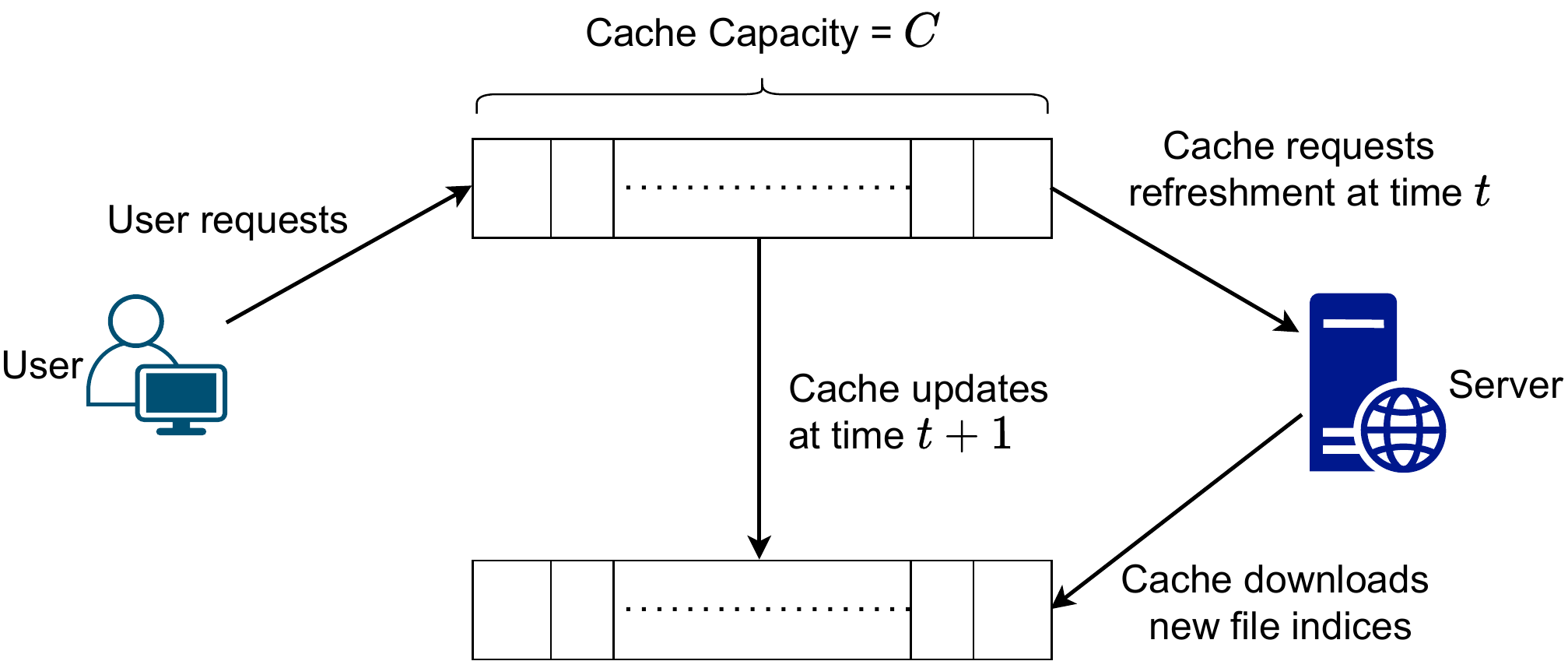}
    \caption{\footnotesize{The online caching set up}}
    \label{fig:cache-illustration}
\end{figure}

In this section, we describe a simplified abstraction of a caching system. 
Assume that a set of $N$ unique files is available in a remote server. A local cache of limited storage capacity can hold up to $C$ files at a time where $C < N$. The system evolves in discrete time-steps, and at each time slot $t$, a  user requests an arbitrary file from the file catalog. The request sequence could be chosen by an oblivious adversary \cite{cesa2006prediction}. An online caching policy $\pi$ determines a set of $C$ files to be cached before the file request for each slot arrives. At any slot $t$, if the requested file is present in the local cache (\emph{i.e.,} in the event of a \emph{cache-hit}), the request is promptly served by the local cache. Otherwise, in the event of a \emph{cache-miss}, the request is forwarded to the remote server, incurring a loss due to the additional routing delay and bandwidth consumption. The caching policy also incurs a switching cost while fetching files into the cache from the remote server. The objective is to design a caching policy that maximizes the caches-hits while incurring minimal switching costs. See Fig.~\ref{fig:cache-illustration} for a schematic. 

Represent the cache configuration at time $t$ by the binary incidence vector $\bm{y}_t\in \{0,1\}^N,$ whose $i$\textsuperscript{th} component $y_{t,i}$ denotes whether the $i^\mathrm{th}$ file is cached at time $t$ or not. The set of all admissible caching configurations $\mathcal{Y}$, which respects the capacity constraint, is given below: 
\begin{align}
    \label{eq:admissible-set}
    \mathcal{Y} & = \left\{\bm{y}\in \{0,1\}^N:\sum_{i\in [N]}y_i\le C\right\}.
\end{align}
Let the variable $f_{t} \in [N]$ denote the user-requested file at time $t$.
Define an associated $N$-dimensional one-hot encoded request vector $\bm{x}_t$ such that $x_{t,i}=\indicator{i=f_t}$. 
\paragraph{Reward/Cost structure}
We combine the reward accrued due to the cache-hits and the fetching cost incurred due to the downloading of files in the cache into a single metric. In particular, we assume that a cache-hit yields unit reward and a cache-miss yields zero reward. Furthermore, we assume that fetching a file into the cache from the remote server costs $D$ units ($D \geq 0$). Hence, changing the cache configuration from $\bm{y}_{t-1}$ to $\bm{y}_t$ at time slot $t$ incurs a total cost of $\frac{D}{2}||\bm{y}_t-\bm{y}_{t-1}||_1.$ Putting everything together, the overall reward obtained at slot $t$ is given by 
 $q_t= \langle \bm{y}_t, \bm{x}_t \rangle - \frac{D}{2}||\bm{y}_t-\bm{y}_{t-1}||_1.$
 
As standard in the online learning literature \cite{cesa2006prediction}, our objective is to design an online caching policy which minimizes the \emph{regret} against an offline static benchmark. Let the vector $\bm{X}_t$ denote the cumulative file request count vector up to time $t$, \emph{i.e.,} $\bm{X}_t = \sum_{\tau=1}^t \bm{x}_{\tau}.$ Taking into account the switching cost as defined above, the expected regret may be expressed as follows:
\begin{align}
\label{eq:regret-formulation}
   \mathbb{E}(R_T) & = \sup_{\bvec{y}\in \mathcal{Y}}\inprod{\bvec{y}}{\bvec{X}_t} - \sum_{t=1}^T \mathbb{E}{\inprod{\bvec{y}_t}{\bvec{x}_t}}\nonumber\\
   \ & +\frac{D}{2}\sum_{t=1}^T \mathbb{E}||\bm{y}_t-\bm{y}_{t-1}||_1,
\end{align}
where the expectation is taken with respect to possible randomness introduced by the caching policy. Note that, the offline static policy does not incur any switching cost as it never changes the cache-configuration. 

%
\section{Performance Analysis}
%
%
\subsection{The \textsf{FTPL} Caching Policy}
The Follow the Perturbed Leader (\textsf{FTPL}) caching policy, introduced in \cite{bhattacharjee2020fundamental}, is described in Algorithm \ref{tab:ftpl-algo}. The policy first samples a set of $N$ i.i.d. standard Gaussian random variables $\bm{\gamma}$. Then, at every slot $t$, it adds a scaled version of the sampled random variables to the current cumulative count vector $\bm{X}_t$ to obtain a perturbed cumulative count vector $\bm{X}_t+\eta_t \bm{\gamma}$. Finally, the \textsf{FTPL} policy caches $C$ files having the highest perturbed cumulative count at slot $t$. The authors in \cite{bhattacharjee2020fundamental} analyzed the expected regret of this policy without considering the switching cost (\emph{i.e.,} the case $D=0$).

\begin{algorithm}[tb!]
\caption{The \textsf{FTPL} Caching Policy}
\label{tab:ftpl-algo}
\begin{algorithmic}[1]
\STATE  Learning rate $\{\eta_t\}_{t \geq 1},$ switching cost $D\geq 0$,  cache capacity $C$, initial cache-configuration $\bm{y}_0$
\STATE  $\bvec{X}_1\gets \bvec{0}$
\STATE \textbf{Sample:} $\bvec{\gamma}\sim \mathcal{N}(\bvec{0},\bvec{I})$.
\FOR {$t=1$ to $T$}
\STATE Cache the top $C$ files corresponding to the perturbed cumulative count vector $\bvec{X}_t+\eta_t \bvec{\gamma}$, \emph{i.e.,} 
\[\bm{y}_t \gets \arg\max_{\bm{y} \in \mathcal{Y}} \langle \bm{y}, \bm{X}_t+ \eta_t \bm{\gamma}\rangle. \]
\label{alg:step-predict}
\STATE User requests a file corresponding to the request vector $\bvec{x}_t$ \label{alg:step-request}
\STATE The policy receives a reward $q_t=\inprod{\bvec{y}_t}{\bvec{x}_t}- \frac{D}{2}\opnorm{\bvec{y}_t - \bvec{y}_{t-1}}{1}$.
\STATE Update $\bvec{X}_{t+1}\gets \bvec{X}_{t} + \bvec{x}_t$.
 \ENDFOR
\end{algorithmic}
\end{algorithm}

\subsection{Analysis of the \textsf{FTPL} policy with Switching Cost}
\label{sec:regret-analysis-ftpl}
%
In Theorem \ref{sw_cost_fixed}, we state and prove the switching regret bound of the \textsf{FTPL} policy for a known time-horizon $T$, using a constant learning rate $\eta_t=\eta, \forall t \geq 1$. In this case, the analysis is simplified by exploiting a property of the \textsf{FTPL} policy, stated in Lemma \ref{lem:atmost-one-switching-per-slot}. However, this property no longer remains true when the learning rate is allowed to vary with time. Hence, the analysis of the \emph{anytime} version of the \textsf{FTPL} policy, which uses time-varying learning rate due to not knowing the time-horizon $T$ a priori, is more subtle and is given in Theorem \ref{thm:anytime-ftpl-switching-regret}. 
\begin{framed}
\begin{thm} \label{sw_cost_fixed}
By choosing the constant learning rate $\eta_t = \eta =\sqrt{T(D+1)/C}(4\pi\ln(N/C))^{-1/4}, \forall t,$ the expected regret of the \textsf{FTPL} caching policy, including the switching cost, is bounded as follows:
\begin{eqnarray*}
 \mathbb{E}(R_T) & \le \frac{2}{\pi^{1/4}}\sqrt{C(D+1)}(\ln(N/C))^{1/4}\sqrt{T},\end{eqnarray*}
 where the expectation is taken w.r.t. the random perturbation $\bm{\gamma}$ added by the policy.
 \end{thm}
\end{framed}
Note that for the case of zero switching cost (\emph{i.e.,} $D=0$), we recover the regret bound of \cite{bhattacharjee2020fundamental}.
\begin{IEEEproof}
We start with the following result for the \textsf{FTPL} policy proved by Bhattacharjee et al. using a stochastically smoothed potential function.
\begin{framed}
\begin{thm}[\cite{bhattacharjee2020fundamental}, Theorem (3)] For the \textsf{FTPL} policy with a constant learning rate $\eta >0,$ we have:
\begin{eqnarray}\label{eq:regret-expression-step1}
\max_{\bvec{y}\in \mathcal{Y}}\inprod{\bvec{y}}{\bvec{X}_t}-\mathbb{E}_{\bm{\gamma}}{\sum_{t=1}^T \inprod{\bvec{y}_t}{\bvec{x}_t}} \nonumber \\ \le  C\eta \sqrt{2\ln (N/C)} + \frac{T}{\eta\sqrt{2\pi}}.
\end{eqnarray}
\end{thm}
\end{framed}
Utilizing the above result, we see that the regret $R_T$ of the \textsf{FTPL} policy with switching cost, as defined in Eqn.\ \eqref{eq:regret-formulation}, is bounded by adding the expected switching cost term:
 \begin{align} \label{reg}
    \ &  \mathbb{E}(R_T) \le C\eta \sqrt{2\ln (N/C)} + \frac{T}{\eta\sqrt{2\pi}} \nonumber\\
    \ & + \frac{D}{2}\sum_{t=2}^T \expectsuff{\bvec{\gamma}}{\opnorm{\bvec{y}_t-\bvec{y}_{t-1}}{1}}.
\end{align}
Hence, it remains to find an upper bound of $\expectsuff{\bvec{\gamma}}{\opnorm{\bvec{y}_t-\bvec{y}_{t-1}}{1}}$ to bound the regret with switching cost in Eqn.\ \eqref{reg}. In this direction, let $S_t$ be the support of the cache configuration vector $\bvec{y}_t$, i.e., \begin{align}
    S_t & = \left\{i\in [N]:y_{t,i}=1\right\}.
\end{align} 
The following lemma shows that under the action of the \textsf{FTPL} policy, the support set $S_t$ can change by only one element at a slot.
\begin{framed}  
\begin{lem}
\label{lem:atmost-one-switching-per-slot}
For any $t\ge 2$, $\abs{S_{t}\setminus S_{t-1}}\le 1$, i.e., at most one entry of the cache can be evicted per slot under the \textsf{FTPL} policy.
\end{lem}
\end{framed}
\textbf{Discussion:} Lemma \ref{lem:atmost-one-switching-per-slot} shows that similar to the classical paging policies, such as, \textsf{LRU}, \textsf{LFU}, and the \textsf{Marker} policy, the proposed \textsf{FTPL} policy also evicts at most one cache element at a time by admitting the currently requested file into the cache. The principal difference among these policies lies in their choice of the file to be evicted from the cache.  
\begin{IEEEproof}
Let the file $f_t$ be requested at time $t$, i.e., $x_{t,f_t}=1$. Thus, we have $X_{t+1,f_t}=X_{t,f_t}+1$, and $X_{t+1,i}=X_{t,i},\ \forall i\ne f_t$. Recall that the set $S_t$ corresponds to the largest $C$ entries of the perturbed cumulative count vector $\bvec{X}_t+\eta \bvec{\gamma}$. Clearly, if $f_t\in S_{t-1}$, then it must be that $S_t=S_{t-1}$. On the other hand, if $f_t\notin S_{t-1}$, $f_t$ may replace at most one file in $S_{t-1}$ to obtain $S_{t-1}$. Hence, it follows that $\abs{S_t\setminus S_{t-1}}\le 1$. 
\end{IEEEproof}
The previous lemma implies that either $\opnorm{\bvec{y}_t-\bvec{y}_{t-1}}{1}=0$, or $\opnorm{\bvec{y}_t-\bvec{y}_{t-1}}{1}=2$. Hence, we have \begin{align}
    \expectsuff{\bvec{\gamma}}{\opnorm{\bvec{y}_t-\bvec{y}_{t-1}}{1}} & = 2\prob{\bvec{y}_t\ne \bvec{y}_{t-1}}.
\end{align}
To get some insight into analyzing the above probability, let us first consider the simplest possible case - a catalog of size two (\emph{i.e.,} $N=2$) and a cache of unit capacity (\emph{i.e.,} $C=1$). Let $g_{t-1}$ denote the index of the only file cached at time $t-1$. Note that, in this case, we have $f_t, g_{t-1} \in \{1,2\}.$ Hence,
\begin{eqnarray}
\label{eq:prelim-expression-case-N=2-C=1}
    &&\mathbb{P}\big(\bvec{y}_{t+1}\ne \bvec{y}_{t}\big)  \nonumber\\
    & = &\mathbb{P}\big(f_t \neq g_{t-1}, X_{t,f_t} + \eta \gamma_{f_t} + 1> X_{t,g_{t-1}} + \eta \gamma_{g_{t-1}}, \nonumber \\
    &&X_{t,f_t} + \eta \gamma_{f_t} \le X_{t,g_{t-1}} + \eta \gamma_{g_{t-1}}\big)\nonumber\\
    & =& \mathbb{P}\big(t-2X_{t,f_t}-1<\sqrt{2}\eta \gamma_{f_t}\le t-2X_{t,f_t}\big),
\end{eqnarray}
where in the last step, we have used the fact that for $f_t \neq g_{t-1},$ we have $X_{t,f_t}+ X_{t, g_{t-1}}=t$. This is true because one file is requested per slot. Since $\gamma_{f_t}$ is a standard Gaussian random variable, from Eqn.\ \eqref{eq:prelim-expression-case-N=2-C=1} we have
\begin{align} \label{eq:ub}
    \lefteqn{\prob{\bvec{y}_{t+1}\ne \bvec{y}_{t}} =  \frac{1}{\sqrt{2\pi}}\int_{\frac{t-1-2X_{t,f_t}}{\sqrt{2}\eta}}^{\frac{t-2X_{t,f_t}}{\sqrt{2}\eta}}e^{-u^2/2}du} & & \nonumber \\
    \ & \leq \frac{1}{\sqrt{2\pi}}\int_{\frac{t-2X_{t,f_t}}{\sqrt{2}\eta}}^{\frac{t-1-2X_{t,f_t}}{\sqrt{2}\eta}} 1 du = \frac{1}{2\eta \sqrt{\pi}}.
\end{align}
We now generalize the above argument to an arbitrary catalog size $N$ and cache capacity $C$. Let $S_{t}$ be the set of cached files corresponding to the cache-configuration $\bvec{y}_t$. We say that a \emph{switching} occurs at time $t$, if the event $\bvec{y}_{t+1}\ne \bvec{y}_t$ takes place, which in turn, is equivalent to the event $\{ f_t\in S_{t+1}$, and $f_t\notin S_t\}$. To compute the probability of this event, let us denote the perturbed cumulative request vector at time $t$ by $\bm{X}'_t,$ \emph{i.e.,} $X'_{t,f}=X_{t,f}+\eta\gamma_f,\ \forall f\in [N]$. Furthermore, let us denote the set of all files \emph{excepting} the $j$\textsuperscript{th} file by $N_j=[N]\setminus \{j\}, \forall j \in [N]$. Sort the components of the vector $\bm{X}'$ in decreasing order and let $X'_{N_j,(f)}$ denote the $f$\textsuperscript{th} component of the sorted vector when we ignore the file $j$ altogether. See Fig.~\ref{fig:cache-selective-sorting} for an illustration.
\begin{figure}[t!]
    \centering
    \includegraphics[height=1.25in,width=3in]{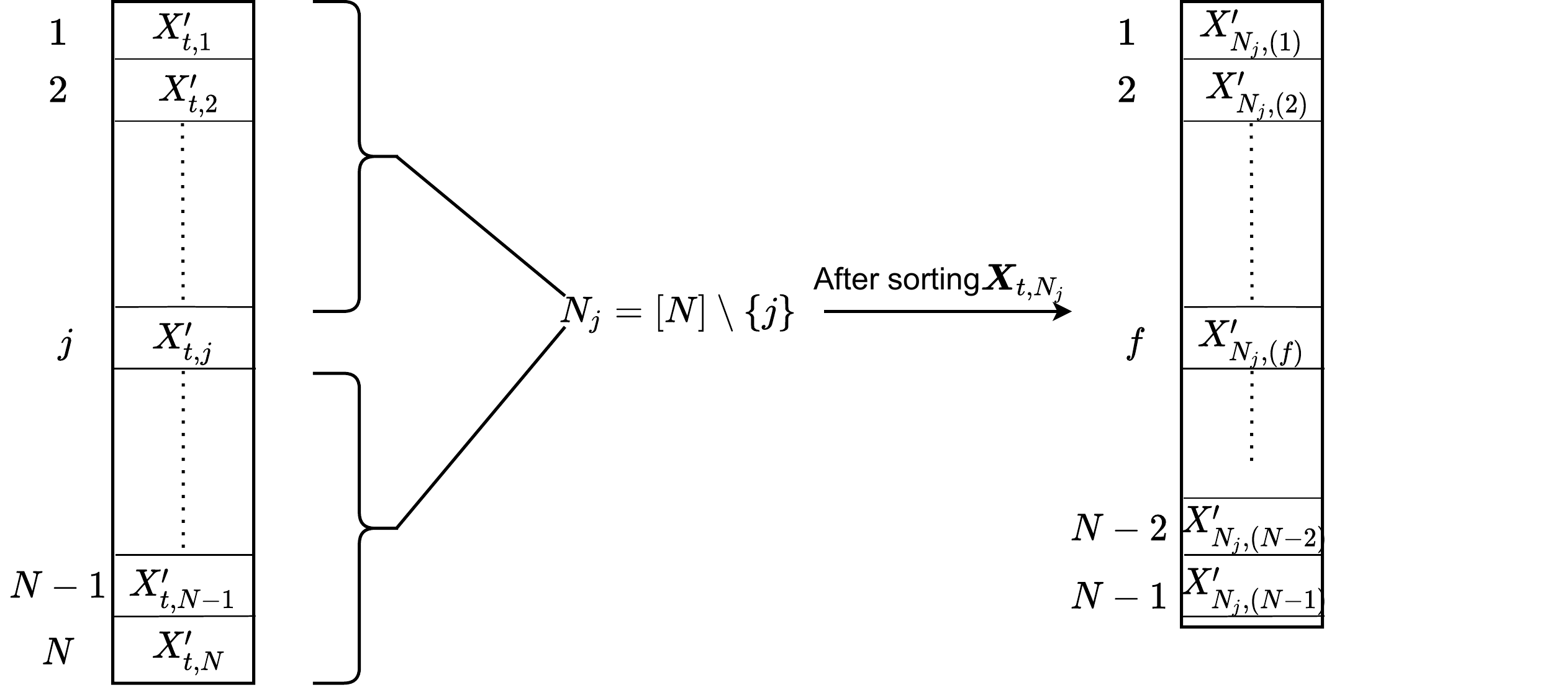}
    \caption{\footnotesize{Illustration of $N_j$ and $X'_{N_j,(f)}$ for some $f\ne j$}.}
    \label{fig:cache-selective-sorting}
\end{figure}
With the above notations, the event $\bvec{y}_{t+1}\ne \bvec{y}_t$ occurs under the \textsf{FTPL} policy if and only if $X'_{N_{f_t},(C)}\ge X'_{t,f_t}$ and $X'_{t,f_t}+1>X'_{N_{f_t},(C)}$. Hence, 
\begin{eqnarray}
\label{eq:switching-probability-exact-base-expression}
   &&\mathbb{P}\big(\bvec{y}_{t+1}\ne \bvec{y}_t\big) \nonumber\\
   &=&\mathbb{P}\big(X'_{N_{f_t},(C)}\ge X'_{t,f_t}>X'_{N_{f_t},(C)}-1\big) \nonumber\\
    &=& \mathbb{P}\big(X'_{N_{f_t},(C)}-X_{t,f_t})/\eta \ge \gamma_{f_t}>\nonumber\\
    &&(X'_{N_{f_t},(C)} -X_{t,f_t})/\eta)-1/\eta\big).
\end{eqnarray}
Note that $f_t$ is deterministic (chosen by the adversary a priori) and, by definition, the random variable $X'_{N_{f_t},(C)}$ does not depend on $\gamma_{f_t}.$
Consequently, conditioning on $\bvec{\gamma}_{N_{f_t}}$, and using the tower-property of conditional expectation, from Eqn.\ \eqref{eq:switching-probability-exact-base-expression} we have:
\begin{eqnarray*}
&&\mathbb{P}(\bm{y}_{t+1} \neq \bm{y}_t)\\
&=& \mathbb{E}\bigg(\mathbb{E}\mathbb{P}\bigg(X'_{N_{f_t},(C)}-X_{t,f_t})/\eta \ge \gamma_{f_t}>\\
&&(X'_{N_{f_t},(C)} -X_{t,f_t})/\eta)-1/\eta|X'_{N_{f_t},(C)} \bigg)
\end{eqnarray*}
Since $\gamma_{f_t}$ is independent of $X'_{N_{f_t}, (C)},$ we can upper bound the inner conditional expectation in the same way as in Eqn. \eqref{eq:ub}, which is independent of $X'_{N_{f_t}, (C)}.$ Taking everything together, we arrive at the same upper bound on switching probability as in Eqn. \eqref{eq:switching-probability-exact-base-expression}:
 \begin{align}
    \prob{\bvec{y}_{t+1}\ne \bvec{y}_t} & \le \frac{1}{\sqrt{2\pi}\eta}.
\end{align}
Combining this result with Eq.~\eqref{eq:regret-expression-step1}, we conclude that the expected regret of the \textsf{FTPL} policy, including the switching cost, is upper-bounded as:
 \begin{align*}
   \mathbb{E}(R_T) & = C\eta \sqrt{2\ln(N/C)} + \frac{T}{\eta\sqrt{2\pi}}+\frac{DT}{\eta\sqrt{2\pi}}\nonumber\\
    \ & =C\eta \sqrt{2\ln(N/C)}+\frac{T}{\eta \sqrt{2\pi}}(1+D).
\end{align*}
Finally, choosing $\eta=\sqrt{T(D+1)/C}(4\pi\ln(N/C))^{-1/4}$, we arrive at the following bound: \begin{align*}
    \mathbb{E}(R_T) & \le \frac{2}{\pi^{1/4}}\sqrt{C(D+1)}(\ln(N/C))^{1/4}\sqrt{T}.
\end{align*}

\end{IEEEproof}

Note that the learning rate in the above caching policy depends explicitly on the horizon length $T$. In the following, we analyze an \emph{anytime} version of the \textsf{FTPL} caching policy with a time-varying learning rate sequence $\{\eta_t\}_{t \geq 1}$ that does not need to know the horizon length $T$ in advance. As mentioned earlier, the analysis of the anytime policy is more involved as Lemma \ref{lem:atmost-one-switching-per-slot}  does not hold in this case.
\section{Analysis of the Anytime \textsf{FTPL} Caching Policy}
\label{sec:regret-analysis-anytime-FTPL}
%
For the anytime version, we use the \textsf{FTPL} policy with a non-decreasing learning rate schedule $\{\eta_t\}_{t\geq 1},$. For pedagogical reasons, the analysis of the switching regret of the anytime \textsf{FTPL} caching policy is divided into several parts. In the following Proposition \ref{prop:anytime_FTPL}, we derive an upper bound on regret \emph{without} the switching cost. This result will later be used for analyzing the regret of the \textsf{FTPL} policy with switching cost. 

\begin{framed}
\begin{proposition} \label{prop:anytime_FTPL}
The regret of the anytime \textsf{FTPL} policy \emph{without} the switching cost (\emph{i.e.,} $D=0$) for any non-decreasing learning rate schedule $\{\eta_t\}_{t\geq 1}$ is given as follows:
\begin{eqnarray*}
&& \hspace{-15pt}\max_{\bvec{y}\in \mathcal{Y}}\inprod{\bvec{y}}{\bvec{X}_t} - \sum_{t=1}^T \expectsuff{\bvec{\gamma}}{\inprod{\bvec{y}_t}{\bvec{x}_t}} \leq  \eta_1C\sqrt{2\log(N/C)}+ \nonumber \\
    && \eta_TC\sqrt{2\ln(Ne/C)} + \frac{1}{\sqrt{2\pi}}\sum_{t=1}^T \frac{1}{\eta_t}.\nonumber
  \end{eqnarray*}
\end{proposition}
\end{framed}
The proof of Proposition \ref{prop:anytime_FTPL} proceeds in the same way as in  \cite{bhattacharjee2020fundamental}, where we additionally take into account the non-constant learning rate schedule $\{\eta_t\}_{t \geq 1}$. See Appendix \ref{prop:anytime_FTPL_proof} for the detailed proof. 

As argued in the previous section, the overall regret of the \textsf{FTPL} policy, including the switching cost, is bounded by the sum of the regret without the switching cost and the expected switching cost (see Eqn.\ \eqref{reg}). Hence, our remaining task is to upper bound the switching cost under the \textsf{FTPL} policy. Towards this goal, we prove the following proposition:
%
\label{sec:switching-cost-anytime-FTPL}
\begin{framed}
\begin{proposition} \label{prop:anytime_FTPL_switching}
The expected switching cost of the \textsf{FTPL} policy with the learning rate schedule $\eta_t = \alpha \sqrt{t} (\alpha>0), t\geq 1$ is bounded as follows:
\begin{align*}
    \lefteqn{\sum_{t=2}^T\expect{\opnorm{\bvec{y}_{t+1}-\bvec{y}_t}{1}} \le \frac{3\sqrt{2}}{\alpha\sqrt{\pi}}\left(\sqrt{T}-1\right)} & &\nonumber\\
    \ & + (N-1)\frac{2+\sqrt{2e\ln(2N)}}{\sqrt{e}}\ln T\nonumber\\
    \ & +\frac{3(N-1)(2+\sqrt{2e\ln(2N)})}{\sqrt{2\pi e}\alpha}\left(1-T^{-1/2}\right).
\end{align*}
\end{proposition}
\end{framed}
\begin{figure*}[t!]
\begin{subfigure}{0.33\textwidth}
    \begin{center}
         \begin{overpic}[height=2in, width=2in]{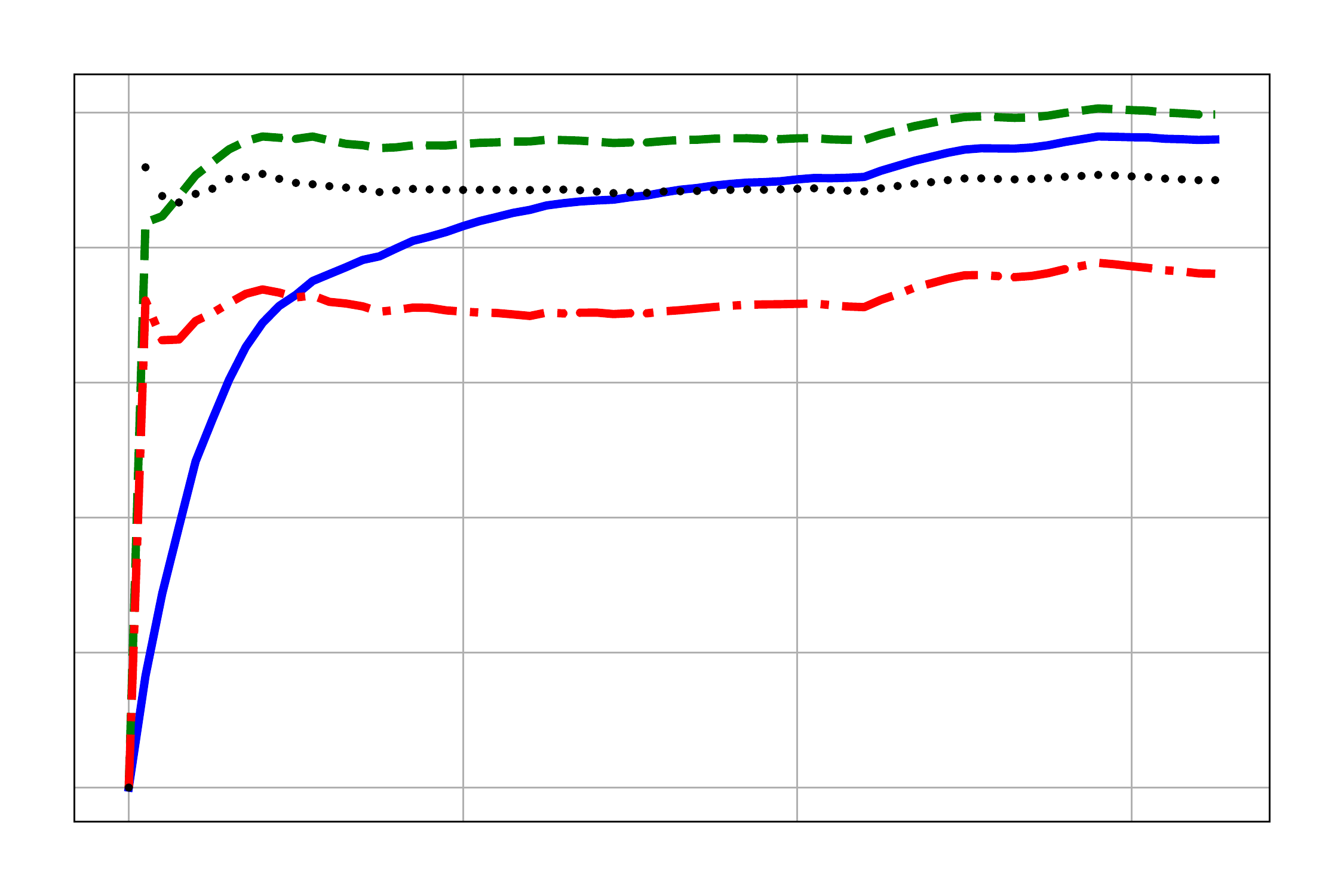}
         \put(13, 100){\tikz \draw[dashed,green, line width = 0.5mm](13,100)--(14,100);\quad \footnotesize{\textsf{FTPL}, $\eta_t=\mathcal{O}(\sqrt{t/C})$}}
         \put(8, 0){\footnotesize{0}}
         \put(33, 0){\footnotesize{2}}
         \put(58, 0){\footnotesize{4}}
         \put(88, 0){\footnotesize{$6\times 10^{4}$}}
         \put(50,-4){$t$}
         \put(-5, 10){\footnotesize{0.0}}
         \put(-5, 40){\footnotesize{0.2}}
         \put(-5, 70){\footnotesize{0.4}}
         \put(-13, 40){\rotatebox{90}{\textrm{\small{Hit rate}}}}
         \end{overpic}
    \end{center}
         \subcaption{}
                  \label{fig:hit-rate-alpha=0.01}
    \end{subfigure}
     \begin{subfigure}{0.33\textwidth}
    \begin{center}
         \begin{overpic}[height=2in, width=2in]{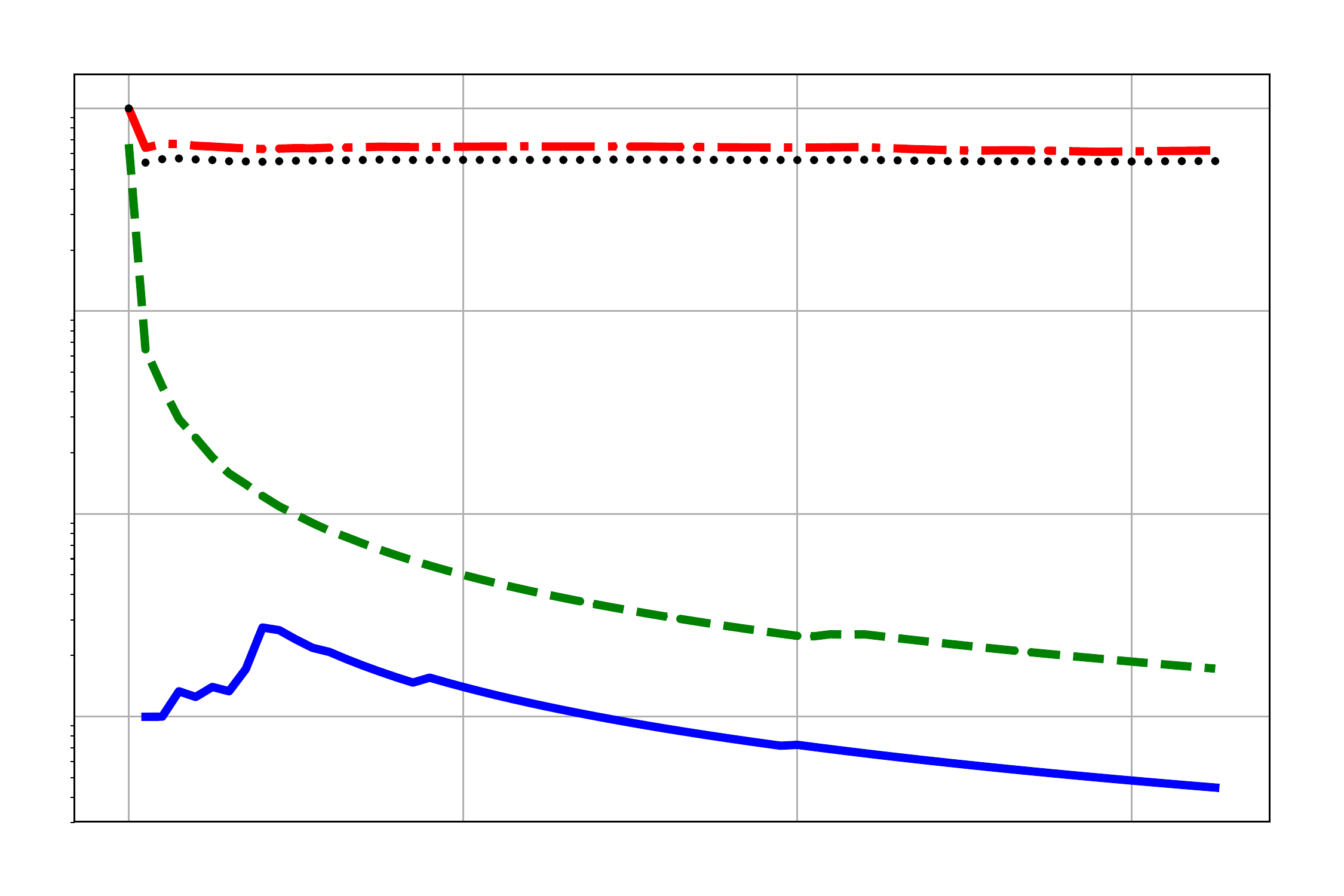}
         \put(3, 100){\tikz \draw[blue, line width = 0.5mm](3,100)--(4,100);\quad \footnotesize{\textsf{FTPL}, fixed $\eta_t$}}
         \put(83, 100){\tikz \draw[dotted,black, line width = 0.5mm](83, 100)--(84,100);\quad \footnotesize{\textsf{LFU}}}
         \put(8, 0){\footnotesize{0}}
         \put(33, 0){\footnotesize{2}}
         \put(58, 0){\footnotesize{4}}
         \put(88, 0){\footnotesize{$6\times 10^{4}$}}
         \put(50,-4){$\footnotesize{t}$}
         \put(-8, 18){\footnotesize{1E-3}}
         \put(-8, 40){\footnotesize{1E-2}}
         \put(-8, 62){\footnotesize{1E-1}}
         \put(-5, 85){\footnotesize{1}}
         \put(-15, 40){\rotatebox{90}{\textrm{\small{Fetch rate}}}}

         \end{overpic}
    \end{center}
         \subcaption{}
              \label{fig:fetch-rate-alpha=0.01}
    \end{subfigure}
    \begin{subfigure}{0.33\textwidth}
    \centering
         \begin{center}
         \begin{overpic}[height=2in, width=2in]{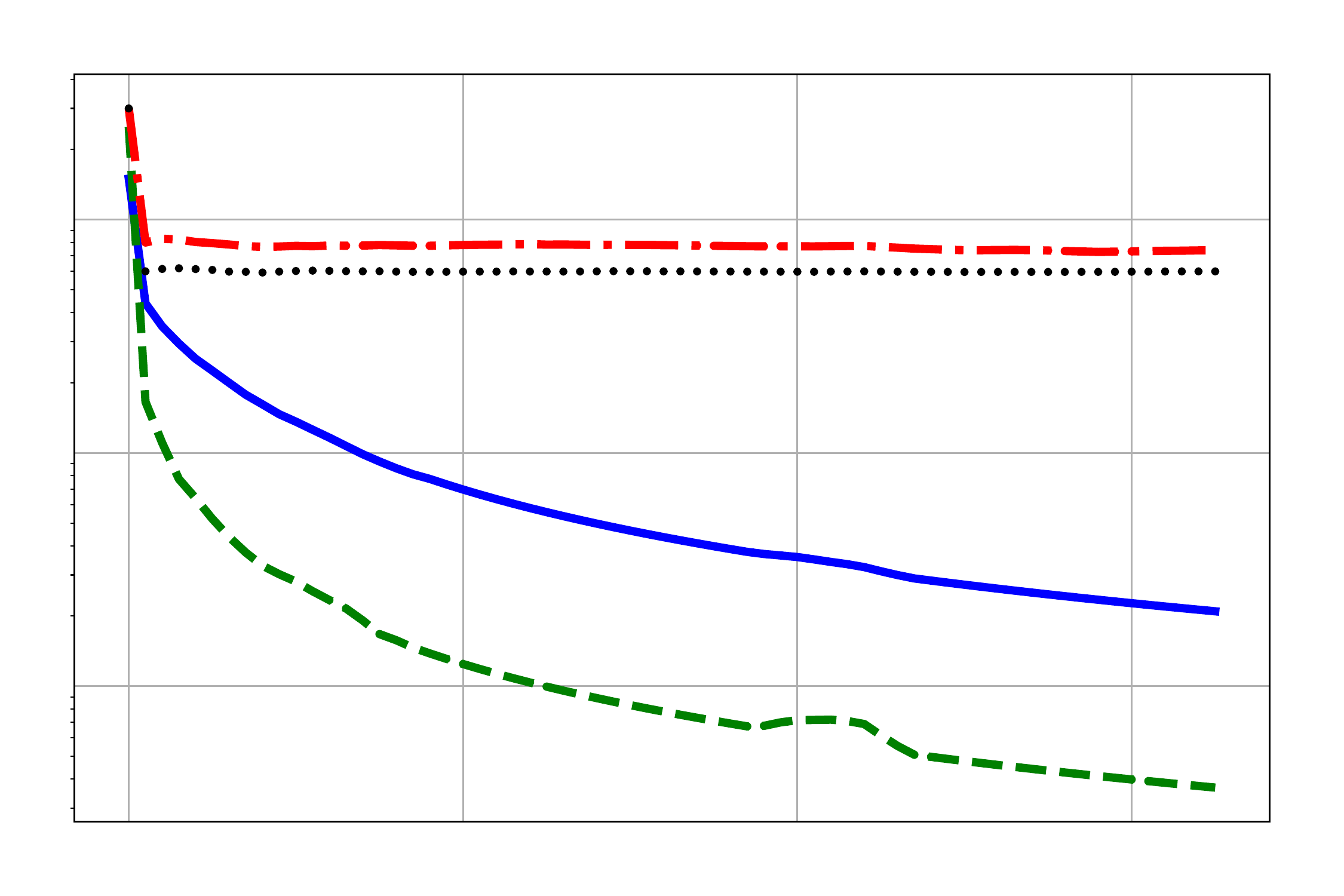}
         \put(13, 100){\tikz \draw[dash dot,red, line width = 0.5mm](13, 100)--(14,100);\quad \footnotesize{\textsf{LRU}}}
         \put(8, 0){\footnotesize{0}}
         \put(33, 0){\footnotesize{2}}
         \put(58, 0){\footnotesize{4}}
         \put(88, 0){\footnotesize{$6\times 10^{4}$}}
         \put(50,-4){$t$}
         \put(-8, 21){\footnotesize{1E-2}}
         \put(-8, 48){\footnotesize{1E-1}}
         \put(-3, 73){\footnotesize{1}}
         \put(-10, 34){\rotatebox{90}{\small{$\dfrac{R_t}{t}$}}}
         \end{overpic}
    \end{center}
         \subcaption{}
                  \label{fig:regret-rate-alpha=0.01}
    \end{subfigure}
    \caption{\small{Comparison of the (a) hit rates, (b) fetch rates, and (c) normalized  regret rates of different caching policies where cache size ($C$) is taken to be $1\%$ of the catalog size ($N$).}}
    \label{fig:cache-comparison-alpha=0.01}
\end{figure*}
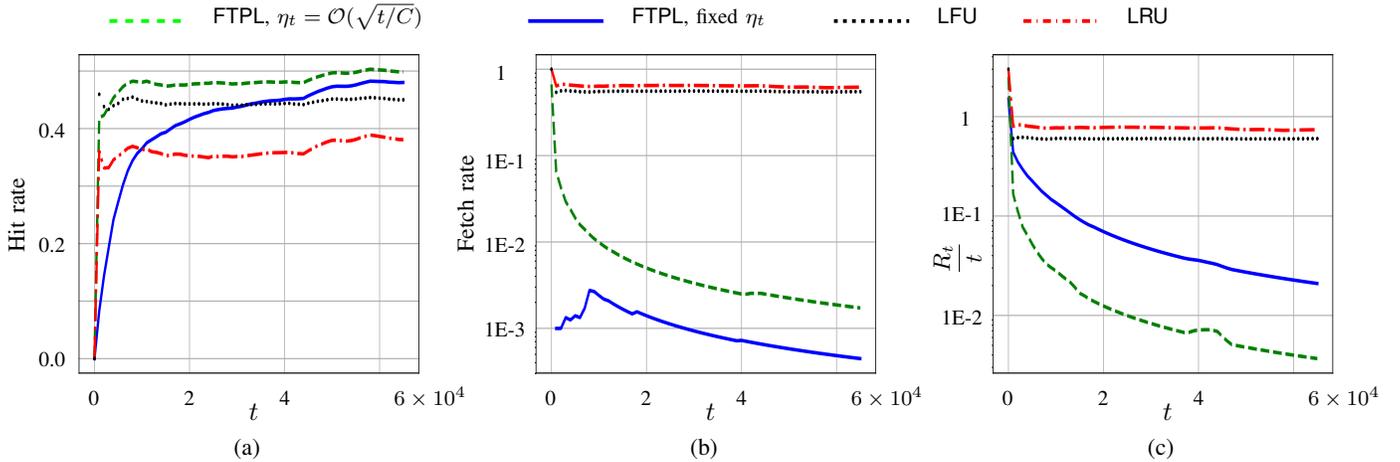
\paragraph{Proof outline} As before, to gain some intuition for the analysis, we consider the simple case $N=2, C=1.$ Then, we carefully bound the probability of new file download at a slot for (a) the requested file and (b) any other files. Note that the latter possibility arises due to the time-varying learning rate schedule. This case does not arise in the case of constant learning-rate considered before. By combining the above two bounds, we obtain the regret bound. 
See Appendix \ref{prop:anytime_FTPL_switching_proof} for the complete proof.

By combining the Propositions~\ref{prop:anytime_FTPL} and~\ref{prop:anytime_FTPL_switching}, we arrive at the following upper bound on the expected regret of the anytime \textsf{FTPL} caching policy with switching cost:
\begin{framed}
    \begin{thm}
    \label{thm:anytime-ftpl-switching-regret}
    With the learning rate schedule $\eta_t=\alpha\sqrt{t}, \alpha>0, t\geq 1$, the expected regret of the anytime \textsf{FTPL} policy, including the switching cost, is bounded as follows:
    \begin{align}
        \label{eq:anytime-ftpl-switching-regret}
        \mathbb{E}(R_T) & \le c_1\sqrt{T} + c_2\ln T + c_3,
    \end{align}
    where $c_1=\mathcal{O}(\sqrt{\ln(Ne/C)})$, and $c_2,c_3$ are small constants depending on $N,C$ and $\alpha$. 
    \end{thm}
\end{framed}
\begin{IEEEproof}
 The proof follows directly from Eq.~\eqref{eq:regret-formulation} using the expression for expected regret and using the upper bounds from Propositions~\ref{prop:anytime_FTPL} and~\ref{prop:anytime_FTPL_switching} along with the simple bound $\sum_{t=2}^T 1/\sqrt{t}\le 2\sqrt{T}-2$.
\end{IEEEproof}

%
%
%
As a direct consequence of Proposition \ref{prop:anytime_FTPL_switching}, it follows that the rate of cache refreshing goes to zero \emph{almost surely} under the action of the \textsf{FTPL} policy. To see this, define the \emph{fetch rate} $\textrm{FR}_t$ at time $t$ to be the total number of file fetches to the cache up to time $t$ divided by $t$. Since the number of switches at a slot is just twice the number of fetches, we have \begin{align}
    \textrm{FR}_t & \stackrel{\Delta}{=} \frac{\sum_{\tau=2}^t\opnorm{\bvec{y}_{\tau}-\bvec{y}_{\tau-1}}{1}}{t}. 
\end{align}
It directly follows from Proposition~\ref{prop:anytime_FTPL_switching} that $\limsup_t\expect{FR_t} = 0$. 
Now, note that for any $t\ge 1$, $FR_t\le C$. Therefore, by the bounded convergence theorem (BCT), we have that $\expect{\limsup_t \textrm{FR}_t} = \limsup_t\expect{\textrm{FR}_t}=0$. Consequently, as $\textrm{FR}_t$ is non-negative, it follows that 
 $\limsup_t \textrm{FR}_t=0$ a.s.
\section{Numerical Experiments}
In our numerical experiments, we use a publicly available anonymized trace from a large commercial CDN server ~\cite{berger2018practical,cdn}. From the trace data, we consider the first $65$K requests, which contain $N=1244$ unique file indices. The cache capacity ($C$) is set to be $1\%$ of the total catalog size ($N$). The plots in Fig.~\ref{fig:cache-comparison-alpha=0.01} compares the performance of the proposed \textsf{FTPL} policy with two other popular caching policies, namely \textsf{LRU}, and \textsf{LFU}. 
From the plot \ref{fig:hit-rate-alpha=0.01}, we see that the proposed \textsf{FTPL} policy out-performs the \textsf{LRU} and the \textsf{LFU} policies in terms of the hit-rate when we consider a relatively large time-horizon ($T\geq 5 \times 10^4$). Furthermore, the next plot \ref{fig:fetch-rate-alpha=0.01} shows that 
the fetch rates of the \textsf{LFU} and the \textsf{LRU} policies are at least two orders of magnitude larger than that of the \textsf{FTPL} policy (for both fixed and time-varying learning rate versions). Finally, Fig. \ref{fig:regret-rate-alpha=0.01} reveals that the normalized regret rate of the \textsf{FTPL} policy diminishes to zero, unlike that of the \textsf{LRU} and \textsf{LFU} policies, which change very little with time.


\section{Conclusion and Future Research}
This paper showed that the \textsf{FTPL}-based caching policy has order-optimal switching regret. The proof involves analyzing the switching cost with Gaussian perturbations and combining the result with the regret bound without switching. Our work improves the best-known switching regret bound for the caching problem by a factor of $O(\sqrt{C})$. Numerical experiments with production-level trace confirm that the proposed \textsf{FTPL} policy out-performs the popular \textsf{LRU} and \textsf{LFU} policies in terms of both hit-rates and download costs. As future work, it will be interesting to extend the results to the case of multiple caches connected to the users in the form of a Bipartite network \cite{bhattacharjee2020fundamental, paria2020caching}. It will also be interesting to obtain bounds on strongly adaptive regret \cite{jun2017improved} which compares the performance of the online algorithm against an offline policy that is allowed to change the cache configuration a fixed number of times.

\clearpage
\bibliography{references}

\begin{thebibliography}{10}
\providecommand{\url}[1]{#1}
\csname url@samestyle\endcsname
\providecommand{\newblock}{\relax}
\providecommand{\bibinfo}[2]{#2}
\providecommand{\BIBentrySTDinterwordspacing}{\spaceskip=0pt\relax}
\providecommand{\BIBentryALTinterwordstretchfactor}{4}
\providecommand{\BIBentryALTinterwordspacing}{\spaceskip=\fontdimen2\font plus
\BIBentryALTinterwordstretchfactor\fontdimen3\font minus
  \fontdimen4\font\relax}
\providecommand{\BIBforeignlanguage}[2]{{%
\expandafter\ifx\csname l@#1\endcsname\relax
\typeout{** WARNING: IEEEtran.bst: No hyphenation pattern has been}%
\typeout{** loaded for the language `#1'. Using the pattern for}%
\typeout{** the default language instead.}%
\else
\language=\csname l@#1\endcsname
\fi
#2}}
\providecommand{\BIBdecl}{\relax}
\BIBdecl

\bibitem{dan1990approximate}
A.~Dan and D.~Towsley, \emph{An approximate analysis of the LRU and FIFO buffer
  replacement schemes}.\hskip 1em plus 0.5em minus 0.4em\relax ACM, 1990,
  vol.~18, no.~1.

\bibitem{lee1999existence}
D.~Lee, J.~Choi, J.-H. Kim, S.~H. Noh, S.~L. Min, Y.~Cho, and C.-S. Kim, ``On
  the existence of a spectrum of policies that subsumes the least recently used
  (lru) and least frequently used (lfu) policies.'' in \emph{SIGMETRICS},
  vol.~99.\hskip 1em plus 0.5em minus 0.4em\relax Citeseer, 1999, pp. 1--4.

\bibitem{pedarsani2016online}
R.~Pedarsani, M.~A. Maddah-Ali, and U.~Niesen, ``Online coded caching,''
  \emph{IEEE/ACM Transactions on Networking (TON)}, vol.~24, no.~2, pp.
  836--845, 2016.

\bibitem{albers}
S.~Albers, \emph{Competitive online algorithms}.\hskip 1em plus 0.5em minus
  0.4em\relax Citeseer, 1996.

\bibitem{cesa2006prediction}
N.~Cesa-Bianchi and G.~Lugosi, \emph{Prediction, learning, and games}.\hskip
  1em plus 0.5em minus 0.4em\relax Cambridge university press, 2006.

\bibitem{bhattacharjee2020fundamental}
R.~Bhattacharjee, S.~Banerjee, and A.~Sinha, ``Fundamental limits on the regret
  of online network-caching,'' in \emph{Abstracts of the 2020
  SIGMETRICS/Performance Joint International Conference on Measurement and
  Modeling of Computer Systems}, 2020, pp. 15--16.

\bibitem{paschos2019learning}
G.~S. Paschos, A.~Destounis, L.~Vigneri, and G.~Iosifidis, ``Learning to cache
  with no regrets,'' in \emph{IEEE INFOCOM 2019-IEEE Conference on Computer
  Communications}.\hskip 1em plus 0.5em minus 0.4em\relax IEEE, 2019, pp.
  235--243.

\bibitem{paria2020caching}
D.~Paria, A.~Sinha \emph{et~al.}, ``Caching in networks without regret,''
  \emph{arXiv preprint arXiv:2009.08228}, 2020.

\bibitem{kalai2005efficient}
A.~Kalai and S.~Vempala, ``Efficient algorithms for online decision problems,''
  \emph{Journal of Computer and System Sciences}, vol.~71, no.~3, pp. 291--307,
  2005.

\bibitem{geulen2010regret}
S.~Geulen, B.~V{\"o}cking, and M.~Winkler, ``Regret minimization for online
  buffering problems using the weighted majority algorithm.'' in
  \emph{COLT}.\hskip 1em plus 0.5em minus 0.4em\relax Citeseer, 2010, pp.
  132--143.

\bibitem{devroye2015random}
L.~Devroye, G.~Lugosi, and G.~Neu, ``Random-walk perturbations for online
  combinatorial optimization,'' \emph{IEEE Transactions on Information Theory},
  vol.~61, no.~7, pp. 4099--4106, 2015.

\bibitem{freund1997decision}
Y.~Freund and R.~E. Schapire, ``A decision-theoretic generalization of on-line
  learning and an application to boosting,'' \emph{Journal of computer and
  system sciences}, vol.~55, no.~1, pp. 119--139, 1997.

\bibitem{littlestone1994weighted}
N.~Littlestone and M.~K. Warmuth, ``The weighted majority algorithm,''
  \emph{Information and computation}, vol. 108, no.~2, pp. 212--261, 1994.

\bibitem{daniely2019competitive}
A.~Daniely and Y.~Mansour, ``Competitive ratio vs regret minimization:
  achieving the best of both worlds,'' in \emph{Algorithmic Learning Theory},
  2019, pp. 333--368.

\bibitem{shanmugam2013femtocaching}
K.~Shanmugam, N.~Golrezaei, A.~G. Dimakis, A.~F. Molisch, and G.~Caire,
  ``Femtocaching: Wireless content delivery through distributed caching
  helpers,'' \emph{IEEE Transactions on Information Theory}, vol.~59, no.~12,
  pp. 8402--8413, 2013.

\bibitem{traverso2013temporal}
S.~Traverso, M.~Ahmed, M.~Garetto, P.~Giaccone, E.~Leonardi, and S.~Niccolini,
  ``Temporal locality in today's content caching: why it matters and how to
  model it,'' \emph{ACM SIGCOMM Computer Communication Review}, vol.~43, no.~5,
  pp. 5--12, 2013.

\bibitem{berger2018practical}
D.~S. Berger, N.~Beckmann, and M.~Harchol-Balter, ``Practical bounds on optimal
  caching with variable object sizes,'' \emph{Proceedings of the ACM on
  Measurement and Analysis of Computing Systems}, vol.~2, no.~2, pp. 1--38,
  2018.

\bibitem{cdn}
``Cdn trace,''
  \url{https://github.com/dasebe/optimalwebcaching/blob/master/README.md},
  accessed: 13-01-2021.

\bibitem{jun2017improved}
K.-S. Jun, F.~Orabona, S.~Wright, and R.~Willett, ``Improved strongly adaptive
  online learning using coin betting,'' in \emph{Artificial Intelligence and
  Statistics}.\hskip 1em plus 0.5em minus 0.4em\relax PMLR, 2017, pp. 943--951.

\bibitem{cohen2015following}
A.~Cohen and T.~Hazan, ``Following the perturbed leader for online structured
  learning,'' in \emph{International Conference on Machine Learning}, 2015, pp.
  1034--1042.

\end{thebibliography}

\clearpage
\section{Appendix} \label{appendix}

\subsection{Proof of Proposition \ref{prop:anytime_FTPL}} \label{prop:anytime_FTPL_proof}
Recall that since the adversary is assumed to be \emph{oblivious}, the file request sequence $\{\bm{x}_t\}_{t \geq 1}$ is chosen before the noise realization $\bm{\gamma}$ is made. Hence, from an analytical point-of-view, the file request sequence is a fixed deterministic sequence which is revealed to the caching policy sequentially in real time.

To analyze this regret, we define (for each time instant $t$) a \emph{time-varying} potential function $\Phi_t:\real^N\to \real$ as below: 
\begin{align}
    \Phi_t(\bvec{x}) & = \expectsuff{\bvec{\gamma}}{\max_{\bvec{y}\in \mathcal{Y}}\inprod{\bvec{y}}{\bvec{x}+\eta_t \bvec{\gamma}}}
\end{align}
Now, note that as $\mathcal{Y}$ is a finite set, for any $\bvec{x}\in \real^N$, $\argmax_{\bvec{y}\in \mathcal{Y}}\expectsuff{\bvec{\gamma}}{\inprod{\bvec{y}}{\bvec{x}+\eta_t\bvec{\gamma}}}$ exists and is unique. Therefore, \begin{align}
    \nabla \Phi_t(\bvec{X}_t) & = \nabla \expectsuff{\bvec{\gamma}}{\inprod{\bvec{y}_t}{\bvec{x}+\eta_t\bvec{\gamma}}}|_{\bvec{x}=\bvec{X}_t} = \expectsuff{\bvec{\gamma}}{\bvec{y}_t}.
\end{align}
Consequently, \begin{align}
    \expectsuff{\bvec{\gamma}}{\inprod{\bvec{y}_t}{\bvec{x}_t}} & = \inprod{\nabla \Phi_t(\bvec{X}_t)}{\bvec{X}_{t+1}-\bvec{X}_t}\nonumber\\
    \ & = \Phi_t(\bvec{X}_{t+1}) - \Phi_t(\bvec{X_t}) - \frac{1}{2}\inprod{\bvec{x}_t}{\nabla^2\Phi_t(\widetilde{\bvec{X}}_t)\bvec{x}_t},
\end{align} where $\widetilde{\bvec{X}}_t=\bvec{X}_t+\theta_t\bvec{x}_t$, for some $\theta_t\in [0,1]$. 
Therefore, \begin{align}
    \lefteqn{\sum_{t=1}^T \expectsuff{\bvec{\gamma}}{\inprod{\bvec{y}_t}{\bvec{x}_t}}} & &\nonumber\\
    \ & = \sum_{t=1}^T \left[\Phi_t(\bvec{X}_{t+1}) - \Phi_t(\bvec{X_t})\right]-\frac{1}{2}\sum_{t=1}^T \inprod{\bvec{x}_t}{\nabla^2\Phi_t(\widetilde{\bvec{X}}_t)\bvec{x}_t}\nonumber\\
    \ & = \Phi_{T}(\bvec{X}_{T+1})-\Phi_1(\bvec{X}_1)+\sum_{t=2}^T \left[\Phi_{t-1}(\bvec{X}_t)-\Phi_t(\bvec{X}_t)\right]\nonumber\\
    \ & -\frac{1}{2}\sum_{t=1}^T \inprod{\bvec{x}_t}{\nabla^2\Phi_t(\widetilde{\bvec{X}}_t)\bvec{x}_t}.
\end{align}
We can follow the same analysis to obtain Eq.(3) in~\cite{cohen2015following}, to obtain a lower bound of the first two terms in the RHS of the equation (the signs will be flipped because of the change in the definition of regret here). Consequently, we obtain, \begin{align}
\label{eq:anytime-regret-analysis-intermediate}
    \lefteqn{\max_{\bvec{y}\in \mathcal{Y}}\inprod{\bvec{y}}{\bvec{X}_t} - \sum_{t=1}^T \expectsuff{\bvec{\gamma}}{\inprod{\bvec{y}_t}{\bvec{x}_t}}} & & \nonumber\\
    \ & \le \Phi_1(\bvec{0}) + \sum_{t=1}^{T-1} \left[\Phi_{t+1}(\bvec{X}_{t+1})-\Phi_{t}(\bvec{X}_{t+1})\right] \nonumber\\
    \ & + \frac{1}{2}\sum_{t=1}^T \inprod{\bvec{x}_t}{\nabla^2\Phi_t(\widetilde{\bvec{X}}_t)\bvec{x}_t}.
\end{align} 
The first term in the RHS is $\Phi_1(\bvec{0})=\eta_1\expectsuff{\bvec{\gamma}}{\max_{\bvec{y}\in \mathcal{Y}}\inprod{\bvec{y}}{\bvec{\gamma}}}$. Using the same analysis as in Theorem 1 of~\cite{cohen2015following}, this can be shown to be upper bounded as \begin{align}
    \Phi_1(\bvec{0}) & \le \eta_1\sqrt{2C\log \abs{\mathcal{Y}}}\nonumber\\
    \ & =\sqrt{2C\log\binom{N}{C}}\le \eta_1C\sqrt{2\log(N/C)},
\end{align}
where we have used the fact that $\abs{\mathcal{Y}}=\binom{N}{C}$.

To analyze the second term in the RHS of the inequality~\eqref{eq:anytime-regret-analysis-intermediate}, we observe that \begin{align}
    \lefteqn{\Phi_{t+1}(\bvec{X}_{t+1})-\Phi_{t}(\bvec{X}_{t+1})} & &\nonumber\\
    \ & =\expectsuff{\bvec{\gamma}}{\max_{\bvec{y}\in \mathcal{Y}}\inprod{\bvec{y}}{\bvec{X}_{t+1}+\eta_{t+1}\bvec{\gamma}} - \max_{\bvec{y}\in \mathcal{Y}}\inprod{\bvec{y}}{\bvec{X}_{t+1}+\eta_{t}\bvec{\gamma}}}\nonumber\\
    \ & \le \expectsuff{\bvec{\gamma}}{\max_{\bvec{y}\in \mathcal{Y}}\inprod{\bvec{y}}{(\eta_{t+1}-\eta_t)\bvec{\gamma}}} = \abs{\eta_{t}-\eta_{t+1}}\expectsuff{\bvec{\gamma}}{\max_{\bvec{y}\in \mathcal{Y}}\inprod{\bvec{y}}{\bvec{\gamma}}}\nonumber\\
    \ & = \abs{\eta_{t+1}-\eta_t}\mathcal{G}(\mathcal{Y}),
\end{align}
where in the second step we have used the inequality $\max_{\bvec{y}\in \mathcal{Y}}(f(\bvec{y})+g(\bvec{y}))\le \max_{\bvec{y}\in \mathcal{Y}}(f(\bvec{y}))+\max_{\bvec{y}\in \mathcal{Y}}(g(\bvec{y}))$, for arbitrary functions $f,g$, and in the last step we have used the definition of Gaussian width of a set.

Finally, the last term in the RHS of inequality~\eqref{eq:anytime-regret-analysis-intermediate} can be upper bounded in exactly the same way as in Eq.(27) of~\cite{bhattacharjee2020fundamental}. Therefore, \begin{align}
    \sum_{t=1}^T \inprod{\bvec{x}_t}{\nabla^2\Phi_t(\widetilde{\bvec{X}}_t)\bvec{x}_t} & \le \sqrt{\frac{2}{\pi}}\sum_{t=1}^T\frac{1}{\eta_t}. 
\end{align}
Putting everything together, and assuming a non-decreasing learning rate schedule $\{\eta_t\}_{t \geq 1}$, we obtain from inequality~\eqref{eq:anytime-regret-analysis-intermediate}, \begin{align}
    \lefteqn{\max_{\bvec{y}\in \mathcal{Y}}\inprod{\bvec{y}}{\bvec{X}_t} - \sum_{t=1}^T \expectsuff{\bvec{\gamma}}{\inprod{\bvec{y}_t}{\bvec{x}_t}}} & & \nonumber\\
    \ & \le \eta_1C\sqrt{2\log(N/C)}+\eta_T\mathcal{G}(\mathcal{Y}) + \frac{1}{\sqrt{2\pi}}\sum_{t=1}^T \frac{1}{\eta_t}\nonumber\\
   \implies R_T & \le \eta_1C\sqrt{2\log(N/C)}+\eta_T\mathcal{G}(\mathcal{Y}) + \frac{1}{\sqrt{2\pi}}\sum_{t=1}^T \frac{1}{\eta_t}\nonumber\\
   \ & + \frac{D}{2}\sum_{t=2}^T\expectsuff{\bvec{\gamma}}{\opnorm{\bvec{y}_t-\bvec{y}_{t-1}}{1}}.
\end{align}

\section{Proof of Proposition \ref{prop:anytime_FTPL_switching}} \label{prop:anytime_FTPL_switching_proof}

One of the main technical difficulties in extending the switching cost analysis of Theorem \ref{sw_cost_fixed} to the anytime case is that Lemma~\ref{lem:atmost-one-switching-per-slot} does not hold here anymore. In fact, due to the time-varying learning rate schedule $\{\eta_t\}_{t \geq 1}$, more than one file may be fetched at a slot. Therefore, a more careful analysis is required for bounding the expected number of switches in the anytime version of the \textsf{FTPL} policy. Throughout the analysis, we will use the following notations.

\paragraph{Notations} Let $E_{ij}(t)$ denote the event that at time $t,$ the $j$\textsuperscript{th} file is fetched into the cache upon evicting the $i$\textsuperscript{th} file. As before, let $f_t$ be the file requested by the adversary at time $t$ and $N_j=[N]\setminus \{j\}$ be the set of all files \emph{excluding} the file $j \in [N]$. 

To gain some insight into the analysis, let use again start with the simplest case of $N=2,C=1$.  In this case, the set $N_{f_t}$ is singleton. Hence, there can be at most one switching at a time $t$. There is the familiar event $E_{N_{f_t},f_t}$, which was analyzed for the fixed $\eta$ case, but for the anytime \textsf{FTPL}, there is also a possibility of the event $E_{f_t,N_{f_t}}$. Therefore, the total switching probability is \begin{align}
    \prob{\bvec{y}_{t+1}\ne \bvec{y}_t} & = \prob{E_{N_{f_t},f_t}} + \prob{E_{f_t,N_{f_t}}}.
\end{align}
Hence, we have
 \begin{align}
    \lefteqn{\prob{E_{N_{f_t},f_t}}} & &\nonumber\\
    \ & = \mathbb{P}\left(X_{t,f_t}+\eta_t\gamma_{f_t}\le X_{t,N_{f_t}}+\eta_t\gamma_{N_{f_t}},\right.\nonumber\\
    \ & \left. X_{t,N_{f_t}}+\eta_{t+1}\gamma_{f_t} + 1>X_{t,N_{f_t}}+\eta_{t+1}\gamma_{N_{f_t}}\right).
\end{align}
Since the requested file $f_t$ is determined by the adversary before the noise $\bm{\gamma}$ is sampled, the r.v.s $\gamma_{f_t}$ and $\gamma_{N_{f_t}}$ are independent. Consequently, we can write that $\gamma_{N_{f_t}}-\gamma_{f_t}=\sqrt{2}Z$, where $Z\sim \mathcal{N}(0,1)$. Hence, defining, $A_{f_t}\equiv X_{t,f_t}-X_{t,N_{f_t}}$, we obtain, \begin{align}
    \prob{E_{N_{f_t},f_t}} & = \prob{\frac{A_{f_t}}{\sqrt{2}\eta_t}\le Z<\frac{A_{f_t}+1}{\sqrt{2}\eta_{t+1}}}\nonumber\\
    \ & \le \frac{1}{2\sqrt{\pi}} \left(\frac{A_{f_t}+1}{\eta_{t+1}}-\frac{A_{f_t}}{\eta_t}\right)_+,
\end{align}
where $x_+ \equiv \max(x,0),\ \forall x\in \real$. Choosing $\eta_t=\alpha \sqrt{t}$ for some $\alpha >0$, we have: \begin{align}
  &  \frac{A_{f_t}+1}{\eta_{t+1}}-\frac{A_{f_t}}{\eta_t}  = \frac{1}{\alpha}\left(\frac{A_{f_t}+1}{\sqrt{t+1}}-\frac{A_{f_t}}{\sqrt{t}}\right).\nonumber
  \end{align}
  Hence, 
  \begin{align}
     & {\alpha}\left(\frac{A_{f_t}+1}{\eta_{t+1}}-\frac{A_{f_t}}{\eta_t}\right) \nonumber \\
    & = A_{f_t}\left(\frac{1}{\sqrt{t+1}}-\frac{1}{\sqrt{t}}\right)+\frac{1}{\sqrt{t+1}}\nonumber\\
    \ & \ge -t\left(\frac{1}{\sqrt{t}}-\frac{1}{\sqrt{t+1}}\right)+\frac{1}{\sqrt{t+1}}\nonumber\\
    \label{eq:A_t-characterization}
    \ & = -\sqrt{t}+\sqrt{t+1}>0.
    \end{align}
    Also, \begin{align}
        \lefteqn{\alpha\left(\frac{A_{f_t}+1}{\eta_{t+1}}-\frac{A_{f_t}}{\eta_t}\right)} & & \nonumber\\
        \ & \le t\left(\frac{1}{\sqrt{t}}-\frac{1}{\sqrt{t+1}}\right)+\frac{1}{\sqrt{t+1}}\nonumber\\
        \ & =\sqrt{\frac{t}{t+1}}\cdot \frac{1}{\sqrt{t}+\sqrt{t+1}}+\frac{1}{\sqrt{t+1}}< \frac{3}{2\sqrt{t+1}}.
    \end{align}
Therefore, with $\eta_t=\alpha \sqrt{t}$, we have, \begin{align}
    \prob{E_{N_{f_t},f_t}} & \le \frac{3}{4\alpha\sqrt{ \pi}\sqrt{t+1}}.
\end{align}

On the other hand, \begin{align}
    \lefteqn{\prob{E_{f_t,N_{f_t}}}} & & \nonumber\\
    \ & = \mathbb{P}\left(X_{t,f_t}+\eta_t\gamma_{f_t}\ge  X_{t,N_{f_t}}+\eta_t\gamma_{N_{f_t}},\right.\nonumber\\
    \ & \left.X_{t,f_t}+\eta_{t+1}\gamma_{f_t} + 1< X_{t,N_{f_t}}+\eta_{t+1}\gamma_{N_{f_t}}\right)\nonumber\\
    \ & = \prob{\frac{A_{f_tt}+1}{\sqrt{2}\eta_{t+1}}<Z\le \frac{A_{f_t}}{\sqrt{2}\eta_t}}\nonumber\\
    \ & \le \frac{1}{2\sqrt{\pi}}\left(\frac{A_{f_t}}{\eta_t}-\frac{A_{f_tt}+1}{\eta_{t+1}}\right)_+=0,
\end{align}
which follows from the characterization in Eq.~\eqref{eq:A_t-characterization}.

Therefore, for $N=2,C=1$, the expected switching cost until time $T$ is upper bounded by $D\sum_{t=1}^{T-1}\frac{3}{8\alpha\sqrt{ \pi}\sqrt{t}}\le \frac{3D}{8\alpha\sqrt{ \pi}}\left(1+2\sqrt{T}\right)$.

We will extend the above analysis of anytime \textsf{FTPL} to the general case with arbitrary $N,C\ (N>C)$. Note that, the number of switches is twice the number of fetches. Therefore, we can write, \begin{align}
    \lefteqn{\expect{\opnorm{\bvec{y}_{t+1}-\bvec{y}_t}{1}}} & & \nonumber\\
    \ & = 2\sum_{f=1}^N \prob{\mbox{The file index $f$ is fetched at time $t+1$}}.
\end{align}
Next, we find an upper bound on the probability that the file $f,\ 1\le f\le N$, is fetched at time $t+1$. 

First, let $f=f_t$. Then, we obtain, \begin{equation}
    \prob{\mbox{$f_t$ is fetched at time $t+1$}} = \prob{f_t\notin S_t, f_t\in S_{t+1}}.
\end{equation}

Before proceeding further, let us define, as in Section~\ref{sec:regret-analysis-ftpl}, for any file $f\in [N]$, $X'_{t,\eta_t,f}=X_{t,f}+\eta_t\gamma_f$. Also, for any $j\in [N]$, sort the components of $X'_{t,\eta_t,f}$ in decreasing orde for $f\in N_{j}$ to re-index them as $X'_{N_j,\eta_t,(1)}\ge \cdots \ge X'_{N_j,\eta_t,(C)}$. 
\subsubsection{Finding an upper bound on $\prob{f_t\in S_{t+1},f_t\notin S_t}$}
\label{sec:upper-bound-ft-fetching}
Using the notation defined before, we have
\begin{align}
    \lefteqn{\prob{f_t\in S_{t+1},f_t\notin S_t}} & &\nonumber\\
    \ & = \mathbb{P}\left(X_{t,f_t}+\eta_t\gamma_{f_t}\le X'_{N_{f_t},\eta_t,(C)},\right.\nonumber\\
    \ & \left.X_{t,f_t}+\eta_{t+1}\gamma_{f_t}+1> X'_{N_{f_t},\eta_{t+1},(C)}\right)\nonumber\\
    \ & = \prob{\frac{X'_{N_{f_t},\eta_{t+1},(C)}-X_{t,f_t}-1}{\eta_{t+1}}<\gamma_{f_t}\le \frac{X'_{N_{f_t},\eta_t,(C)}-X_{t,f_t}}{\eta_t}}\nonumber\\
    \label{eq:ft-switching-probabilityintermediate-expression}
    \ & \le \frac{1}{\sqrt{2\pi}}\mathbb{E}_{\bvec{\gamma}_{N_{f_t}}}\left[\left(\frac{1}{\eta_{t+1}}+X_{t,f_t}\left(\frac{1}{\eta_{t+1}}-\frac{1}{\eta_t}\right)+\frac{X'_{N_{f_t},\eta_t,(C)}}{\eta_t}\right.\right.\nonumber\\
    \ & \left.\left.-\frac{X'_{N_{f_t},\eta_{t+1},(C)}}{\eta_{t+1}}\right)_+\right],
\end{align}
where we have used the fact that since $f_t$ is independent of the randomness in the algorithm, so that the random variable $\gamma_{f_t}$ is independent of $\bvec{\gamma}_{N_{f_t}}$.
Now, note that, for any $f\in [N]$, \begin{align}
    \frac{X'_{t,\eta_{t+1},f}}{\eta_{t+1}} = \frac{X_{t,f}}{\eta_{t+1}}+\gamma_{f}=\frac{X'_{t,\eta_t,f}}{\eta_t} + X_{t,f}\left(\frac{1}{\eta_{t+1}}-\frac{1}{\eta_t}\right).
\end{align}
Since, for any $f\in N_{f_t}$, $0\le X_{t,f}\le t-X_{t,f_t}$, it follows from the last inequality that, for any $f\in N_{f_t}$, 
\begin{align}
\label{eq:observation}
    0 & \le \frac{X'_{t,\eta_t,f}}{\eta_t} - \frac{X'_{t,\eta_{t+1},f}}{\eta_{t+1}} \le (t-X_{t,f_t})\left(\frac{1}{\eta_{t}}-\frac{1}{\eta_{t+1}}\right).
\end{align}
We now claim the following:
\begin{framed}
\begin{lem}
\label{lem:ordering-lemma}
For any positive integer $n$, if $a_1,\cdots a_n$ and $b_1,\cdots, b_n$ are real numbers such that $a_i\le b_i,\ 1\le i\le n$, then $a_{(j)}\le b_{(j)}$ for $1\le j\le n$, where $a_{(j)}(b_{(j)})$ are the ordered values of $a_i(b_i)$ such that $a_{(1)}\ge \cdots a_{(n)}\ (b_{(1)}\ge \cdots \ge b_{(n)})$.
\end{lem}
\end{framed}
\begin{IEEEproof}
 To prove the result, we first observe that there exists a permutation $\pi$ of the indices $[n]$ such that $a_{\pi(1)}=a_{(1)},\cdots, a_{\pi(n)}=a_{(n)}$. Then, for any $1\le j\le n$, note that $a_{(j)}\le a_{(l)}=a_{\pi(l)}\le b_{\pi(l)}$ for all $1\le l\le j-1$, and that $a_{(j)}=a_{\pi(j)}\le b_{\pi(j)}$. Therefore, $a_{(j)}\le \min\{b_{\pi(1)},\cdots, b_{\pi(j)}\}\le b_{(j)}$. This proves the claim.  
\end{IEEEproof}
Now, let us take $a_i = \frac{X'_{t,\eta_{t+1},i}}{\eta_{t+1}},\ b_i=\frac{X'_{t,\eta_t,i}}{\eta_{t}},\ i\in N_{f_t}$. Then from the left inequality of~\eqref{eq:observation}, it follows using Lemma~\ref{lem:ordering-lemma} that \begin{align}
    \frac{X'_{N_{f_t},\eta_{t+1},(C)}}{\eta_{t+1}} & \le \frac{X'_{N_{f_t},\eta_{t},(C)}}{\eta_{t}}. 
\end{align}
Similarly, taking $a_i = \frac{X'_{t,\eta_t,i}}{\eta_{t}}$, and $b_i=\frac{X'_{t,\eta_{t+1},i}}{\eta_{t+1}} + (t-X_{t,f_t})\left(\frac{1}{\eta_{t}}-\frac{1}{\eta_{t+1}}\right)$, it follows from Lemma~\ref{lem:ordering-lemma} that 
\begin{align}
    \frac{X'_{N_{f_t},\eta_{t},(C)}}{\eta_{t}} & \le \frac{X'_{N_{f_t},\eta_{t+1},(C)}}{\eta_{t+1}} + (t-X_{t,f_t})\left(\frac{1}{\eta_{t}}-\frac{1}{\eta_{t+1}}\right). 
\end{align} 
Therefore, the expression inside the first brackets in the RHS of inequality~\eqref{eq:ft-switching-probabilityintermediate-expression} is lower bounded by $\frac{1}{\eta_{t+1}}+X_{t,f_t}\left(\frac{1}{\eta_{t+1}}-\frac{1}{\eta_{t}}\right)$. This quantity is lower bounded by $\frac{1}{\eta_{t+1}}-t\left(\frac{1}{\eta_{t}}-\frac{1}{\eta_{t+1}}\right)$, which was previously shown to be non-negative. On the other hand, the RHS is upper bounded by  
\begin{align}
    \frac{1}{\eta_{t+1}}+\left(t-2X_{t,f_t}\right)\left(\frac{1}{\eta_{t}}-\frac{1}{\eta_{t+1}}\right) & \le \frac{1}{\eta_{t+1}}+t\left(\frac{1}{\eta_{t}}-\frac{1}{\eta_{t+1}}\right)\nonumber\\
    \ & \le \frac{3}{2\alpha \sqrt{t+1}}.
\end{align} 
Therefore, we obtain, \begin{align}
    \prob{f_t\in S_{t+1} , f_t\notin S_t} & \le \frac{3}{2\alpha \sqrt{2\pi}\sqrt{t+1}}.
\end{align}
%
%
\subsubsection{Upper-bounding $\mathbb{P\big(}$\textrm{~The ~file~} $f$ \textrm{~is ~fetched ~at ~} $t+1\big)$ \textrm{~for ~some~} $f\ne f_t$}
To upper-bound this probability, let $\Omega_f(t)$ denote the event that the file $f$ is fetched at time instant $t$. Then, using the law of total probability, we can write: 
\begin{align}
\lefteqn{\prob{\Omega_f(t+1)}}& &  \nonumber\\
\label{eq:prob-f-switching-decomposition}
\ & = \prob{\Omega_{f}(t+1)\cap\Omega_{f_t}(t+1)} + \prob{\Omega_{f}(t+1)\cap \overline{\Omega}_{f_t}(t+1)}.
\end{align} 
Next, we separately upper-bound the above two probabilities. For any set $S\in [N]$, define the set $N_S=[N]\setminus S$. Let $F_t := \{f,f_t\}$, and assume that $N-2\ge C$ in the rest of the analysis.
\paragraph{Upper-bounding $\prob{\Omega_{f}(t+1)\cap\Omega_{f_t}(t+1)}$}
Recall that the event $\Omega_f(t+1)\cap \Omega_{f_t}(t+1)$ is the event that the files $f$ and $f_t$ were not in the set $S_t$ but are in $S_{t+1}$. Therefore, 
\begin{align}
  \lefteqn{ \prob{\Omega_f(t+1)\cap \Omega_{f_t}(t+1)} = \prob{F_t \subset S_{t+1}\setminus S_t}} & &\nonumber\\
  \ & = \mathbb{P}\left(\max_{j\in F_t}X'_{t,\eta_t,j} \le X'_{N_{F_t},\eta_t,(C)},\right.\nonumber\\
  \ & \left. \min\{X'_{t,\eta_{t+1},f},X'_{t,\eta_{t+1},f_t}+1\}> X'_{N_{F_t},\eta_{t+1},(C)}\right)\nonumber\\
   \ & = \mathbb{P}\left(\{X'_{t,\eta_t,f}\le X'_{N_{F_t},\eta_t,(C)},X'_{t,\eta_{t+1},f}> X'_{N_{F_t},\eta_{t+1},(C)}\}\right.\nonumber\\
   \ & \left.\cap \{X'_{t,\eta_t,f_t}\le X'_{N_{F_t},\eta_t,(C)},X'_{t,\eta_{t+1},f_t}+1> X'_{N_{F_t},\eta_{t+1},(C)}\}\right).
  \end{align}
Note that, by definition, the random variables $X'_{N_{F_t},\eta_t,(C)}$ and $X'_{N_{F_t},\eta_{t+1},(C)}$ are almost surely determined by the random variables $\gamma_{j}, j\in F_t$. Furthermore, $F_t=\{f,f_t\}$ is a fixed, non-random set of indices, where $f$ is a fixed chosen index, and $f_t$ is an index chosen by the adversary. Since the adversary is oblivious, $f_t$ is independent of the random variables $\gamma_1,\cdots, \gamma_{N_{F_t}}$. Consequently, because of the mutual independence of the random variables $\gamma_1,\cdots, \gamma_N$, we can conclude that the random variables $X_{t,\eta_t,j},\ j\in F_t$ and $X'_{N_{F_t},\eta_t,(C)}$ are mutually independent. Similar conclusion holds true for the set of random variables $X_{t,\eta_{t+1},j},\ j\in F_t$ and $X'_{N_{F_t},\eta_{t+1},(C)}$. Therefore, one can write, 
\begin{align}
    \lefteqn{\prob{\Omega_f(t+1)\cap \Omega_{f_t}(t+1)}} & &\nonumber\\
    \ & = \mathbb{E}_{\bvec{\gamma}_{N_{F_t}}}\left[\mathbb{P}\left(X'_{t,\eta_t,f_t}\le X'_{N_{F_t},\eta_t,(C)},\right.\right.\nonumber\\
    \ & \left.\left.X'_{t,\eta_{t+1},f_t}> X'_{N_{F_t},\eta_{t+1},(C)}-1\right)\right.\cdot\nonumber\\
    \ & \left.\prob{X'_{t,\eta_t,f}\le X'_{N_{F_t},\eta_t,(C)},X'_{t,\eta_{t+1},f}> X'_{N_{F_t},\eta_{t+1},(C)}}\right].
\end{align}    
Now, given $\bvec{\gamma}_{N_{F_t}}$, we obtain, as in Section~\ref{sec:upper-bound-ft-fetching},
\begin{align}
    \lefteqn{\prob{X'_{t,\eta_t,f_t}\le X'_{N_{F_t},\eta_t,(C)},X'_{t,\eta_{t+1},f_t}> X'_{N_{F_t},\eta_{t+1},(C)}-1}} & &\nonumber\\
    \ & = \prob{\frac{X'_{N_{F_t},\eta_{t+1},(C)}-1-X_{t,f_t}}{\eta_{t+1}}<\gamma_{f_t}\le \frac{X'_{N_{F_t},\eta_t,(C)}-X_{t,f_t}}{\eta_t}}\nonumber\\
    \ & \le \frac{1}{\sqrt{2\pi}}\left(\frac{X'_{N_{F_t},\eta_t,(C)}-X_{t,f_t}}{\eta_t} - \frac{X'_{N_{F_t},\eta_{t+1},(C)}-1-X_{t,f_t}}{\eta_{t+1}}\right)_+\nonumber\\
    \ & \le \frac{1}{\sqrt{2\pi}}\left((t-X_{t,f}-2X_{t,f_t})\left(\frac{1}{\eta_t}-\frac{1}{\eta_{t+1}}\right)+\frac{1}{\eta_{t+1}}\right)\nonumber\\
    \ & \le \frac{1}{\sqrt{2\pi}}\left(t\left(\frac{1}{\eta_t}-\frac{1}{\eta_{t+1}}\right)+\frac{1}{\eta_{t+1}}\right)\nonumber\\
    \ & < \frac{3}{2\alpha\sqrt{2\pi(t+1)}}.
\end{align}
Now, given $\bvec{\gamma}_{N_{F_t}}$, let us consider the event $X'_{t,\eta_t,f}\le X'_{N_{F_t},\eta_t,(C)},X'_{t,\eta_{t+1},f}> X'_{N_{F_t},\eta_{t+1},(C)}$, which in words, is equivalent to the event that $X'_{t,\eta_t,f}$ is smaller than the $C^\mathrm{th}$ largest element of $X'_{t,\eta_t,j},\ j\in N_{F_t}$, while $X'_{t,\eta_{t+1},f}$ is larger than the $C^{\mathrm{th}}$ largest element of $X'_{t,\eta_{t+1},j},\ j\in N_{F_t}$. Let us call the set of the top $C$ elements of $X'_{t,\eta_t,j}, j\in N_{F_t}$ as $\Lambda_{t}$. Evidently, this event implies that there must be an index $i\in \Lambda_t$ such that $X'_{t,\eta_t,f}\le X'_{t,\eta_t,i}$ and $X'_{t,\eta_{t+1},f}>X'_{t,\eta_{t+1},i}$. To verify this, first observe that it follows from the event $X'_{t,\eta_t,f}\le X'_{N_{F_t},\eta_t,(C)}$, that for all $i\in \Lambda_t$, $X'_{t,\eta_t,f}\le X'_{t,\eta_t,i}$. Now, if $X'_{t,\eta_{t+1},f}\le X'_{t,\eta_{t+1},i}$ for all $i\in \Lambda_t$, then obviously, one cannot have $X'_{t,\eta_{t+1},f}> X'_{N_{F_t},\eta_{t+1},(C)}$, as $\Lambda_t\subset N_{F_t}$.

Now, consider sorting $X'_{t,\eta_{t+1},j},\ j\in N_{F_t}$ in decreasing order. Define 
\begin{align}
    i_t = \min\{j\in \Lambda_t: X'_{t,\eta_{t+1},(C)}\ge X'_{t,\eta_{t+1},j}\}, 
\end{align}
where we assume that the indices in $\Lambda_t$ are sorted in increasing order, i.e., from smallest to largest. It follows from the pigeonhole principle that $i_t$ exists. Consequently, we obtain that the event  $X'_{t,\eta_t,f}\le X'_{N_{F_t},\eta_t,(C)},X'_{t,\eta_{t+1},f}> X'_{N_{F_t},\eta_{t+1},(C)}$ implies that $X'_{t,\eta_t,f}<X'_{t,\eta_t,i_t}$ and $X'_{t,\eta_{t+1},f}>X'_{t,\eta_{t+1},i_t}$. Note that the index $i_t$ is \emph{entirely determined} by $\bvec{\gamma}_{N_{F_t}}$ and therefore is independent of $\gamma_f$.  

Therefore, one can write that, given $\bvec{\gamma}_{N_{F_t}}$, 
\begin{align}
    \lefteqn{\prob{X'_{t,\eta_t,f}\le X'_{N_{F_t},\eta_t,(C)},X'_{t,\eta_{t+1},f}> X'_{N_{F_t},\eta_{t+1},(C)}}} & & \nonumber\\
    \ & \le \prob{X'_{t,\eta_t,f}\le X'_{t,\eta_t,i_t},X'_{t,\eta_{t+1},f}> X'_{t,\eta_{t+1},i_t}}\nonumber\\
    \ & =\prob{\frac{X_{t,i_t}-X_{t,f}}{\eta_{t+1}}<\gamma_{f}-\gamma_{i_t}\le \frac{X_{t,i_t}-X_{t,f}}{\eta_{t}}}.
\end{align}
Note that the above probability can be non-zero only when $X_{t,i_t}>X_{t,f}$, since other wise, we have $\frac{X_{t,i_t}-X_{t,f}}{\eta_{t}}<\frac{X_{t,i_t}-X_{t,f}}{\eta_{t+1}}$. Therefore, we obtain, given $\bvec{\gamma}_{N_{F_t}}$, 
\begin{align}
    \lefteqn{\prob{\frac{X_{t,i_t}-X_{t,f}}{\eta_{t+1}}<\gamma_{f}-\gamma_{i_t}\le \frac{X_{t,i_t}-X_{t,f}}{\eta_{t}}}} & & \nonumber\\
    \ & = \max\left\{0, \int_{\frac{X_{t,i_t}-X_{t,f}}{\eta_{t+1}}}^{\frac{X_{t,i_t}-X_{t,f}}{\eta_{t}}}\frac{e^{-(\gamma+\gamma_{i_t})^2/2}}{\sqrt{2\pi}}d\gamma\right\}.
\end{align}
Now, note that, for any $0\le a<b$, 
\begin{align}
\lefteqn{\int_a^b e^{-(\gamma+\gamma_{i_t})^2/2}d\gamma} & &\nonumber\\
\ & \le (b-a)\left[e^{-(a+\gamma_{i_t})^2/2}\indicator{\gamma_{i_t}>-a}+e^{-(b+\gamma_{i_t})^2/2}\indicator{\gamma_{i_t}<-b}\right.\nonumber\\
\ & \left.+\indicator{-b\le \gamma_{i_t}\le -a}\right].    
\end{align} 
The three terms in the right hand side above can be upper bounded as below:
\begin{align}
    \lefteqn{(b-a)e^{-(a+\gamma_{i_t})^2/2}} & &\nonumber\\
    \ & =(b/a-1)(a+\gamma_{i_t})e^{-(a+\gamma_{i_t})^2/2}\nonumber\\
    \ & -(b/a-1)\gamma_{i_t}e^{-(a+\gamma_{i_t})^2/2}\nonumber\\
    \ & \le \frac{(b/a-1)}{\sqrt{e}} +(b/a-1)\abs{\gamma_{i_t}}, 
\end{align}
where we have used the inequality $xe^{-x^2/2}\le e^{-1/2},\ \forall x\in \real$. Similarly, 
\begin{align}
(b-a)e^{-(b+\gamma_{i_t})^2/2} \le \frac{(1-a/b)}{\sqrt{e}} + (1-a/b)\abs{\gamma_{i_t}}.    
\end{align}
Now, note that we can write
\begin{align}
\lefteqn{(b-a)\indicator{-b\le \gamma_{i_t}\le -a}} & &\nonumber\\
\ & \le (b/a-1)(-\gamma_{i_t}) \indicator{-b\le \gamma_{i_t}\le -a}\nonumber\\
\ & \le (b/a-1)\abs{\gamma_{i_t}}\indicator{-b\le \gamma_{i_t}\le -a}.   
\end{align}
Putting everything together, we obtain, 
\begin{align}
    \lefteqn{\int_a^b e^{-(\gamma+\gamma_{i_t})^2/2}dt} & & \nonumber\\
    \ & \le \left(\frac{(b/a-1)}{\sqrt{e}} +(b/a-1)\abs{\gamma_{i_t}}\right)\indicator{\gamma_{i_t}>-a}\nonumber\\
    \ & + \left(\frac{(1-a/b)}{\sqrt{e}} + (1-a/b)\abs{\gamma_{i_t}}\right)\indicator{\gamma_{i_t}<-b} +\nonumber\\
    \ & + (b/a-1)\abs{\gamma_{i_t}}\indicator{-b\le \gamma_{i_t}\le -a}\nonumber\\
    \ & = \frac{(b-a)}{\sqrt{e}}\left(\frac{\indicator{\gamma_{i_t}>-a}}{a}+\frac{\indicator{\gamma_{i_t}<-b}}{b}\right)\nonumber\\
    \ & + (b/a-1)\abs{\gamma_{i_t}}+ (2-a/b-b/a)\abs{\gamma_{i_t}}\indicator{\gamma_{i_t}<-b}\nonumber\\
    \ & \le \frac{b^2-a^2}{\sqrt{e}ab}+(b/a-1)\abs{\gamma_{i_t}},
\end{align}
where in the last step we have used $\indicator{\gamma_{i_t}<-b},\indicator{\gamma_{i_t}>-a}\le 1$ and the fact that $b/a-1\ge 1-a/b$ as $b, a\ge 0$.

Consequently, 
\begin{align}
   \lefteqn{ \prob{\frac{X_{t,i_t}-X_{t,f}}{\eta_{t+1}}<\gamma_{f}-\gamma_{i_t}\le \frac{X_{t,i_t}-X_{t,f}}{\eta_{t}}}} & & \nonumber\\
   \ & \le \left(\sqrt{\frac{t+1}{t}}-\sqrt{\frac{t}{t+1}}+\sqrt{e}\left(\sqrt{\frac{t+1}{t}}-1\right)\abs{\gamma_{i_t}}\right)\frac{1}{\sqrt{e}}\nonumber\\
   \label{eq:prob-f-switching-with-N-Ft}
    \ & < \frac{1}{t\sqrt{e}}+\frac{\abs{\gamma_{i_t}}}{2t}.
\end{align}
 Therefore, we obtain, 
 \begin{align}
     \lefteqn{\prob{\Omega_f(t+1)\cap \Omega_{f_t}(t+1)}} & & \nonumber\\
     \ & < \expectsuff{\bvec{\gamma}_{N_{F_t}}}{\frac{3}{2\alpha\sqrt{2\pi(t+1)}}\cdot \left(\frac{1}{t\sqrt{e}}+\frac{\abs{\gamma_{i_t}}}{2t}\right)}\nonumber\\
     \ & < \frac{3}{2\alpha\sqrt{2\pi}t^{3/2}}\left(\frac{1}{\sqrt{e}}+\frac{\expect{\abs{\gamma_{i_t}}}}{2}\right).
 \end{align}
 Now, noting that $\abs{\gamma_{i_t}}\le \max_{i\in [N]}\abs{\gamma_i}$, we obtain, $\expect{\abs{\gamma_{i_t}}}\le \expect{\max_{i\in [N]}\abs{\gamma_i}}\le \sqrt{2\ln(2N)}$, where the last step uses Massart's Lemma for Gaussians. Therefore, 
 \begin{align}
     \lefteqn{\prob{\Omega_f(t+1)\cap \Omega_{f_t}(t+1)}} & &\nonumber\\
     \label{eq:prob-f-ft-switch}
     \ & < \frac{3}{4\alpha\sqrt{2\pi e}t^{3/2}}\left(2+\sqrt{2e\ln(2N)}\right)=\frac{\mathcal{O}(\sqrt{\ln N})}{t^{3/2}}.
 \end{align}
\paragraph{Finding an upper bound of $\prob{\Omega_f(t+1)\cap \overline{\Omega}_{f_t}(t+1)}$} 
First note that the event $\Omega_f(t+1)\cap \overline{\Omega}_{f_t}(t+1)$ is equivalent to the event that $f$ is fetched and $f_t$ is not fetched. The event that $f_t$ is not fetched is equivalent to either that $f_t$ does not switch from $t$ to $t+1$, or due to the event that $f_t$ was in the cache at time $t$ and is evicted at time $t+1$. Now, when $f_t$ does not switch, it immediately implies that the switching of $f$ occurs with one of the top $C$ files in $N_{F_t}$. Therefore, we can write, 
\begin{align}
    \lefteqn{\prob{\Omega_f(t+1)\cap \overline{\Omega}_{f_t}(t+1)}} & & \nonumber\\
    \ & \le \prob{X'_{t,\eta_t,f}\le X'_{N_{F_t},\eta_t,(C)},X'_{t,\eta_{t+1},f}> X'_{N_{F_t},\eta_{t+1},(C)}}\nonumber\\
    \ & +\prob{\Omega_f(t+1)\cap \{\mbox{$f_t$ is evicted from $S_t$}\}}\nonumber\\
    \ & \le \prob{X'_{t,\eta_t,f}\le X'_{N_{F_t},\eta_t,(C)},X'_{t,\eta_{t+1},f}> X'_{N_{F_t},\eta_{t+1},(C)}}\nonumber\\
    \ & +\prob{\mbox{$f_t$ is evicted from $S_t$}}.
\end{align}
Now, note that the probability of evicting $f_t$ can be evaluated as below: 
\begin{align}
    \lefteqn{\prob{\mbox{$f_t$ is evicted from $S_t$ at instant $t+1$}}} & & \nonumber\\
    \ & = \prob{X'_{t,\eta_t,f_t}\ge X'_{N_{f_t},\eta_t,(C)}, X'_{t,\eta_{t+1},f_t}+ 1<X'_{N_{f_t},\eta_{t+1},(C)}}\nonumber\\
    \ & = \prob{\frac{X'_{N_{f_t},\eta_t,(C)}-X_{t,f_t}}{\eta_t}\le \gamma_{f_t}< \frac{X'_{N_{f_t},\eta_{t+1},(C)}-1-X_{t,f_t}}{\eta_{t+1}}}.
\end{align}
But note that 
\begin{align}
\lefteqn{\frac{X'_{N_{f_t},\eta_t,(C)}-X_{t,f_t}}{\eta_t} - \frac{X'_{N_{f_t},\eta_{t+1},(C)}-1-X_{t,f_t}}{\eta_{t+1}}} & &\nonumber\\
\ & \ge X_{t,f_t}\left(\frac{1}{\eta_t}-\frac{1}{\eta_{t+1}}\right)+\frac{1}{\eta_{t+1}}\nonumber\\
\ & \ge -t\left(\frac{1}{\eta_t}-\frac{1}{\eta_{t+1}}\right) + \frac{1}{\eta_{t+1}}\nonumber\\
\ & =\frac{1}{\alpha}\left(-t\frac{1}{\sqrt{t}\sqrt{t+1}(\sqrt{t}+\sqrt{t+1})}+\frac{1}{\sqrt{t+1}}\right)\nonumber\\
\ & \ge \frac{1}{\alpha}\left(-\frac{1}{2\sqrt{t+1}}+\frac{1}{\sqrt{t+1}}\right)>0.
\end{align}
Therefore, \begin{align}
    \prob{\mbox{$f_t$ is evicted from $S_t$ at instant $t+1$}} = 0.
\end{align}
Consequently, \begin{align}
    \lefteqn{\prob{\Omega_f(t+1)\cap \overline{\Omega}_{f_t}(t+1)}} & & \nonumber\\
    \ & \le  \prob{X'_{t,\eta_t,f}\le X'_{N_{F_t},\eta_t,(C)},X'_{t,\eta_{t+1},f}> X'_{N_{F_t},\eta_{t+1},(C)}}\nonumber\\
    \ & \le \expectsuff{\bvec{\gamma}_{N_{F_t}}}{\frac{1}{t}\left(\frac{1}{\sqrt{e}}+\frac{\abs{\gamma_{i_t}}}{2}\right)}\nonumber\\
    \label{eq:prob-f-switching-without-ft}
    \ & \le \frac{1}{t}\left(\frac{1}{\sqrt{e}}+\sqrt{\frac{\ln(2N)}{2}}\right),
\end{align}
where, in the penultimate step, we have used the result in Eq.~\eqref{eq:prob-f-switching-with-N-Ft} and in the last step, we have used the Massart's Lemma for Gaussians.

Therefore, taking together the inequalities.~\eqref{eq:prob-f-switching-without-ft},~\eqref{eq:prob-f-ft-switch} and using Eq.~\eqref{eq:prob-f-switching-decomposition}, we obtain, 
\begin{align}
    \lefteqn{\prob{\mbox{$f$ is fetched at time $t+1$}}} & & \nonumber\\
    \ & \le \frac{1}{t}\left(\frac{1}{\sqrt{e}}+\sqrt{\frac{\ln(2N)}{2}}\right)\nonumber\\
    \ & + \frac{3}{4\alpha\sqrt{2\pi e}t^{3/2}}\left(2+\sqrt{2e\ln(2N)}\right)\nonumber\\
    \ & = \frac{\sqrt{2}+\sqrt{e\ln(2N)}}{\sqrt{2e}}\left(\frac{1}{t}+\frac{3}{2\sqrt{2\pi}\alpha t^{3/2}}\right).
\end{align}
Taking everything together, we obtain, 
\begin{align}
    \lefteqn{\sum_{t=2}^T\expect{\opnorm{\bvec{y}_{t+1}-\bvec{y}_t}{1}}} & & \nonumber\\
    \ & \le \frac{3}{\alpha\sqrt{2\pi}}\sum_{t=2}^T \frac{1}{\sqrt{t}} \nonumber\\
    \ & + \sqrt{2}(N-1)\frac{\sqrt{2}+\sqrt{e\ln(2N)}}{\sqrt{e}}\sum_{t=2}^T \left(\frac{1}{t}+\frac{3}{2\sqrt{2\pi}\alpha t^{3/2}}\right).
\end{align}
Now, we use the standard inequalities, 
\begin{align}
    \sum_{t=2}^T \frac{1}{\sqrt{t}} & \le \int_1^T \frac{1}{\sqrt{t}} dt = 2\sqrt{T}-2,\\
    \sum_{t=2}^T \frac{1}{t} & \le \int_1^T \frac{1}{t}dt=\ln T,\\
    \sum_{t=2}^T \frac{1}{t^{3/2}} & \le \int_1^T \frac{1}{t^{3/2}} = 2 - \frac{2}{\sqrt{T}},
\end{align}
to obtain, 
\begin{align}
    \lefteqn{\sum_{t=2}^T\expect{\opnorm{\bvec{y}_{t+1}-\bvec{y}_t}{1}}} & & \nonumber\\
    \ & \le \frac{3\sqrt{2}}{\alpha\sqrt{\pi}}\left(\sqrt{T}-1\right) \nonumber\\
    \ & + \sqrt{2}(N-1)\frac{\sqrt{2}+\sqrt{e\ln(2N)}}{\sqrt{e}}\left(\ln T +\frac{3}{\sqrt{2\pi}\alpha}\left(1-\frac{1}{\sqrt{T}}\right)\right)\nonumber.
\end{align}

\end{document}